\documentclass[a4paper,11pt]{article}
\usepackage{jheppub} 
\usepackage{graphicx,dcolumn,bm,epsfig,cancel,hyperref}
\usepackage{amsmath}
\usepackage{subcaption}
\usepackage{amssymb, times}

\newcommand{\comment}[1]{{}}

\newcommand{\hl}{\color{black}}

\usepackage{fontawesome}

\makeatletter
\newcommand{\doublewidetilde}[1]{{%
  \mathpalette\double@widetilde{#1}%
}}
\newcommand{\double@widetilde}[2]{%
  \sbox\z@{$\m@th#1\widetilde{#2}$}%
  \ht\z@=.9\ht\z@
  \widetilde{\box\z@}%
}
\makeatother
\usepackage{color}
\title{
  Two-Scalar Bose-Einstein Condensates: From Stars to Galaxies
}
\author[a]{Huai-Ke Guo}
\author[a]{Kuver Sinha}
\author[b]{Chen Sun}
\author[c]{Joshua Swaim}
\author[a]{Daniel Vagie}
\affiliation[a]{Department of Physics and Astronomy, University of Oklahoma, Norman, OK 73019, USA}
\affiliation[b]{School of Physics and Astronomy, Tel-Aviv University, Tel-Aviv 69978, Israel}
\affiliation[c]{Vassar College, Dept. of Physics \& Astronomy, Box 745, Poughkeepsie, NY 12604}
\emailAdd{ghk@ou.edu}
\emailAdd{kuver.sinha@ou.edu}
\emailAdd{chensun@mail.tau.ac.il}
\emailAdd{jswaim@vassar.edu}
\emailAdd{Daniel.d.vagie-1@ou.edu}
\begin{document}

%\date{\today}

\abstract { We study the properties of Bose-Einstein Condensate (BEC)
  systems consisting of two scalars, focusing on both the case where
  the BEC is stellar scale as well as the case when it is galactic
  scale. After studying the stability of such systems and making
  contact with existing single scalar limits, we undertake a numerical
  study of the two interacting scalars using Einstein-Klein-Gordon
  (EKG) equations, including both non-gravitational self-interactions
  and interactions between the species. We show that the presence of extra
  scalars and possible interactions between them can leave unique
  imprints on the BEC system mass profile, especially when the system
  transitions from being dominated by one scalar to being dominated by
  the other.  At stellar scales (nonlinear regime,) we observe that a
  repulsive interaction between the two scalars of the type
  $+\phi_1^2 \phi_2^2$ can stabilize the BEC system and support it up
  to high compactness, a phenomenon only known to exist in the
  $+\phi^4$ system. We provide simple analytic understanding of this
  behavior and point out that it can lead to interesting gravitational
  wave signals at LIGO-Virgo.  At galactic scales, on the other hand,
  we show that two-scalar BECs can address the scaling problem that
  arises when one uses ultralight dark matter mass profiles to fit observed
  galactic core mass profiles. In the end, we construct a particle
  model of two ultralight scalars with the repulsive
  $\phi_1^2 \phi_2^2$ interaction using collective symmetry
  breaking. We develop a fast numerical code that utilizes the
  relaxation method to solve the EKG system, which can be easily
  generalized to multiple
  scalars. \href{https://github.com/vagiedd/BosonStars}{\faGithub} }

\maketitle
\date{\today}
%\pacs{xxx.xxx}
%\preprint{xxx-xxx}
%\tableofcontents
%

\section{Introduction}
\label{sec:introduction}

In the last few years, there has been increasing interest in ultra-light bosonic dark matter (DM)  candidates such as the axion. While the QCD axion was originally motivated by the strong CP problem \cite{Weinberg:1977ma,Wilczek:1977pj}, string theory  predicts a vast landscape of axion-like particles (ALPs) \cite{Svrcek:2006yi,Arvanitaki:2009fg,Cicoli:2012sz,Acharya:2010zx} with masses across several orders of magnitude and a rich phenomenology. Studies of sub-eV (pseudo-) scalars as DM  candidates have yielded interesting signals and novel proposals for direct detection experiments \cite{JacksonKimball:2017elr,Garcon:2019inh,Ouellet:2018beu,Bloch:2019lcy,Graham:2020kai}, to name a few.

In particular, due to its bosonic nature, ultra-light bosonic DM can exhibit collective behaviors at the macroscopic level that are not obvious at the Lagrangian level. It has been observed and well understood in condensed matter physics that for bosons there exists a unique phase, the Bose-Einstein Condensate (BEC) phase, once the ensemble is cooled down below the critical temperature. Given the abundance of the DM population, this translates the requirement of the occupancy number $n > (mv)^3$ to an upper bound of the scalar mass, $m < \mathrm{eV}$ \cite{Fan:2016rda}.
The maximal mass of the BEC object can be crudely estimated as $M\lesssim M_{Pl}^2/m$ \cite{Colpi:1986ye}. This singles out two scales of particular interest to the community: galactic scale BEC with $m \sim 10^{-22} \;\mathrm{eV}$, and stellar scale BEC with $m \sim 10^{-10} \;\mathrm{eV}$.

On galactic scales, condensates of ultralight bosons have been shown to produce core like halos by quantum pressure \cite{Hui:2016ltb,Schive:2014dra,Schive:2014hza,Schwabe:2016rze,Veltmaat:2016rxo,Mocz:2017wlg}, with various studies on its constraints \cite{Amorisco:2018dcn,10.1093/mnras/stx1870,Schutz:2020jox,Bar:2018acw,Deng:2018jjz}. A good understanding of the theoretical mass profile of such a BEC system not only provides insights on the particle nature of DM, but could also have implications for quasar lensing time delay and the recent Hubble tension \cite{Blum:2020mgu}. On smaller scales, such BEC systems can form stellar scale structures dubbed boson stars~\cite{Giudice:2016zpa,Liebling:2012fv}, with scalars free from interaction \cite{1995PhDT........25G}, with attractive $\phi^4$ interaction \cite{Schiappacasse:2017ham,Chavanis:2011zi,Chavanis:2011zm,Visinelli:2017ooc,Eby:2018ufi,Eby:2017teq,Eby:2018dat,Eby:2015hyx,Eby:2016cnq}, repulsive $\phi^4$ interaction \cite{Colpi:1986ye,PhysRevD.38.2376,Hertzberg:2020xdn}, and repulsive $\cos(\phi/f)$ potential \cite{Fan:2016rda,Croon:2018ybs}. A few variations such as multistate boson stars from  generic scalars have also been explored, in an attempt to reproduce realistic models of DM halos \cite{Bernal_2010}. BEC states with angular momentum is studied in a recent work \cite{Kling:2020xjj}. % In a recent study \cite{Eby:2020eas}, it is shown that it can be extended to two flavor axion states which have noticible distinctions in the structure compared to a single scalar field. 
 New ways of probing BEC systems at different scales include using  Big Bang Nucleosynthesis (BBN) \cite{Blum:2014vsa}, galaxy rotation curves  \cite{Bar:2019bqz,Bar:2018acw}, gravitational wave (GW) from binary boson star mergers \cite{Bezares:2018qwa,Croon:2018ybs}, {\hl GW from BEC collisions \cite{Helfer:2018vtq}, speed of GWs passing through BEC \cite{Dev:2016hxv},} electromagnetic emission \cite{Hertzberg:2018zte,Hertzberg:2020dbk,Amin:2020vja}, GW from extreme mass ratio inspiral systems \cite{Guo:2019sns}, and optical lensing \cite{Prabhu:2020pzm}.

% The collective behavior of BEC systems leads to emergent phenomena that are not obvious at the Lagrangian level.
Many studies have been dedicated to understanding the map between properties at the Lagrangian level and the behavior of the BEC system such as its mass and density profile \cite{Schiappacasse:2017ham,Deng:2018jjz,Eby:2017teq,Eby:2018dat,Eby:2015hyx,Eby:2016cnq,Croon:2018ybs,Fan:2016rda,Croon:2018ftb,PhysRevD.38.2376,Berezhiani:2015bqa,Chavanis:2011zi,Chavanis:2011zm,Ferreira:2018wup}. {\hl In \cite{Chavanis:2011cz}, hydrodynamic approach is used and confirms the results from field space analysis. In \cite{Chen:2020cef}, formation of boson stars inside DM halo is simulated.} On the other hand, the effect of extra scalars in a BEC system with both gravitational and possibly non-gravitational interactions among the scalars remains largely under-explored. In particular, due to numerical challenges, most previous studies have focused on single scalar BEC systems, with a few exceptions: multiple scalar BEC systems with negligible non-gravitational interactions were explored in \cite{Broadhurst:2018fei};  analytical approximations for multi-scalar BEC systems with self-interactions were explored in \cite{Eby:2020eas}; {\hl a Newtonian analysis on multi-scalar BEC in the limit of large quartic coupling \cite{Kan:2017uhj}. Existence of solutions in the presence of a few types of interactions are studied in \cite{Brihaye:2007tn,Brihaye:2008cg,Brihaye:2009yr}.} In contrast, we undertake a full {\hl General Relativity (GR)} numerical study of the properties of BECs made of two interacting scalars, including both non-gravitational self-interactions and interactions between the scalars, followed by simple analytical understanding, {\hl and its phenomenological implications.} Given that feeble repulsive self-interactions can lead to drastic changes in the mass profile at the macroscopic level \cite{Colpi:1986ye,Croon:2018ybs}, it can be expected that interactions between different scalars will have an important impact and leave unique imprints on the BEC system mass profile. %The purpose of this paper is to initiate a systematic numerical study of the properties of BECs made of \textit{multiple interacting scalars}, including both non-gravitational self-interactions and interactions between the scalars. 
The purpose of our paper is to carefully investigate such imprints, and ask whether they can be utilized to predict unique observational signatures or help address long-standing puzzles.

 Our study proceeds along two directions. At the stellar scale, a light scalar of mass $m\sim 10^{-10} \;\mathrm{eV}$ allows the formation of solar mass stellar structures. The formation and compactness can be greatly enhanced due to the presence of a repulsive self-interaction in the scalar potential, or compromised by an attractive self-interaction \cite{Giudice:2016zpa,Schiappacasse:2017ham,Croon:2018ybs}. The strength and form of the non-gravitational interactions leave imprints on the GW signal. With the presence of extra scalars and interactions between multiple scalars, the features in GW are richer, with the maximal compactness of a stable BEC system being $\sim \mathcal{O}(0.2)$. {\hl In particular, we show the important role the interaction between the species ($\pm \phi_1^2 \phi_2^2$) plays in either stabilizing or destabilizing the self-gravitating two scalar BEC system.} % {\color{red} model-wise... signal-wise... formation history wise (merger of axion stars)...}
 It could also have implications for the recent GW190521 event. With this perspective in mind, we explore the mass versus compactness parameter space of a stellar BEC consisting of two ultralight scalars.

The second major focus of our paper is the behavior of the BEC at galactic scales. In the case of BEC DM composed of a single scalar field, the mass profile has a unique scaling behavior controlled by a single parameter: the central value of the wave function ($\propto$ (central density)$^{1/2}$ of the BEC.) As shown in \cite{Bar:2018acw,Deng:2018jjz}, this scaling behavior is in tension with observational data \cite{Rodrigues:2017vto}. {\hl This can be understood from the scaling behavior of the scalar's equation of motion, either from the Schr\"odinger-Newton equation, or from the relativistic EKG equation: in the single scalar case, the scaling is parametrized as $M[\phi(0)], R[\phi(0)]$, with $\phi(0)$ being the central value of the classical wave function. This means that dynamics does not play any role in determining the mass-radius relation, $M(R)$, which is fixed once a scalar potential is chosen}. With the presence of a second scalar, the theory space is enlarged from one dimensional $\{\phi(0)\}$ to two dimensional  $\{\phi_1(0), \phi_2(0)\}$. {\hl As we will show in subsequent sections, the ratio of the two BEC structures plays a role in the mass-radius relation of the total BEC structure, meanwhile the system is stable against radial perturbation even the fraction of each component varies.} This holds out the possibility of accommodating observational data with BEC DM composed of multiple scalars {\hl while maintaining stability against radial perturbation}; we show that this is indeed the case. {\hl We stress that, we are not claiming to provide a mass-radius relation of the BEC, as that requires accounting for the dynamics of the scalars that determines the mass ratio of the two components, which is beyond the scope of this work. Instead, we study the stability of the BEC against radial perturbations, when a ratio of the two components is given. Under this assumption, we show that there is parameter space to accommodate the galactic data \cite{Rodrigues:2017vto} and address the problem raised by \cite{Deng:2018jjz}. As the first part of a study series, we lay out the ground work and justifies the necessities of studying multi-scalar ultralight dark matter dynamics. }

 Our results rely on numerically integrating the complete relativistic Einstein-Klein-Gordon (EKG) equations. We do so by implementing an efficient algorithm not common in boson star studies to solve the set of equation of motion for arbitrary parameters.  Our algorithm employs the Relaxation Method described in \cite{NumericalRecipes} which can solve a system of differential equations subjected to their boundary conditions as opposed to the initial value shooting method typically used in these type of equations. This approach can be easily extended to N-scalar systems.\footnote{We note that the relaxation method avoids the problem of multi-dimension shooting, yet it still suffers from the stiffness issue if the separation of scales is large. That means, in the case of $N$-scalars, the separation between the lightest one and the heaviest one cannot be too large.} The results are verified by comparing against the usual shooting method in the single scalar case \cite{Croon:2018ybs,Guo:2019sns}. Besides the exact numerical solutions, we also provide analysis adopting a simple ansatz and write out the non-relativistic Hamiltonian. The scaling behavior in the linear regime is affected mainly by the  mass ratio ($m_1/m_2$) of the two scalars, while that in the nonlinear regime is affected more by the non-gravitational interactions between the two scalars. In particular, we demonstrate that with mild repulsive interactions between the two scalars $\lambda_{12} \phi_1^2 \phi_2^2$, the system can be stabilized up to very large denstiy, a behavior that was only known to exist in the case of repulsive self-interaction $\lambda \phi^4$ \cite{Colpi:1986ye,PhysRevD.38.2376,Schiappacasse:2017ham,Croon:2018ybs}. We show that such a repulsive interaction can be realized in a realistic particle model with collective symmetry breaking \cite{Low:2002ws}.

To summarize, we highlight the following points in this work:
\begin{itemize}
  \item At the galactic scale, we show that the presence of a second scalar renders the theory capable of accommodating the mass profile indicated by observational data {\hl while maintaining its radial stability,} which cannot be done in a system of a single scalar \cite{Deng:2018jjz}.
  \item At the stellar scale, we show that a repulsive interaction between two scalars, $\lambda \phi_1^2 \phi_2^2$, can stabilize the system up to high density, which was only known to exist in a single scalar system with repulsive self-interactions \cite{Colpi:1986ye,PhysRevD.38.2376,Croon:2018ybs,Guo:2019sns}. We provide a particle model realization as a proof of concept.
  
  \item We have developed complete and fast code that utilizes the Relaxation Method to solve the BEC system with two scalars, and is easy to generalize to multiple scalars. The code has been made public.\footnote{The code can be downloaded at https://github.com/vagiedd/BosonStars .} 
%  \item To verify the stability of such BEC systems with two scalars, we perform numerical radial perturbations, and check the time evolution explicitly. 
\end{itemize}

This paper is organized as follows. We will begin by defining the phenomenological model of two scalars in Section~\ref{sec:bose-einst-cond}. We set up the stage for numerical computations of the Einstein-Klein-Gordon system. We then take the non-relativistic limit to simplify the system and perform analytical investigations of the behavior, including both the transition from one scalar dominating to the other, and the effect of the non-gravitational interactions between the two scalars. We then verify the static solutions numerically, as well as perform time evolution of the system to ensure that the solution is indeed stable against radial perturbations. In Section~\ref{sec:galactic-scale-bec} we apply our analysis toolkit to the galactic scale BEC system  and show that this can address the scaling problem of ultralight dark matter
{\hl while maintaining radial stability.} % studied in \cite{Deng:2018jjz}.
In Section~\ref{sec:BosonStars} we focus on the stellar scale and show the effect of non-gravitational interactions in the context of two scalar system. In Section~\ref{sec:model-realization} we discuss a possible particle model construction. We then conclude in Section~\ref{sec:Summary}.
We provide details for computing the equation of motion in Appendix~\ref{sec:Einstein-tensor}, outline the numerical recipe we use for the code in Appendix~\ref{sec:NumericalProcedure}, and verify that it can reproduce the single scalar limit in Appendix~\ref{sec:single-scalar-limit}.

\section{Bose-Einstein Condensate with Multiple Scalars}
\label{sec:bose-einst-cond}

\subsection{Phenomenological Model} 
The Lagrangian for a complex scalar system consisting of $N$ particles reads

\begin{equation}
    \mathcal{L} = \sum_{n}^N -\frac{1}{2} g^{\mu \nu} \partial_\mu \phi^{*}_i \partial_\nu \phi_i- V(|\phi_{1}|^2,|\phi_{2}|^2,...)
\end{equation}
where $g^{\mu \nu}$ is the space-time metric inverse with the signature $(-,+,+,+)$, and $\phi_{n}$ is the $n$-th scalar field.  The potential $V$ characterizes the interactions between the scalar fields and is a function of the coupling strengths and the modulus squared of $\phi$. % We note that system can either describe N scalars all in the same state or N states in separate states~\cite{Bernal_2010}. 
In this work we only consider the case of two complex scalars in the ground state, which can be easily generalized to compute BEC of more scalar fields in our numerical framework. The Lagrangian for 2 complex scalars with generic interactions reads 
\begin{align}
  \label{eq:Lagrangian-KG-curved-space}
  \mathcal{L} =
    &-\frac{1}{2} g^{\mu \nu}\partial_\mu \phi_1^* \partial_\nu \phi_1 -
    \frac{1}{2} m_1^2 |\phi_1|^2 
    - \frac{1}{4}c_1|\phi_1|^4 \cr
    &-\frac{1}{2} g^{\mu \nu}\partial_\mu \phi_2^* \partial_\nu \phi_2 -
    \frac{1}{2} m_2^2 |\phi_2|^2 
    - \frac{1}{4}c_2|\phi_2|^4
    - \frac{1}{4} c_{12} |\phi_1|^2|\phi_2|^2,
\end{align}
where $c's$ are the coupling constants that can be either positive or negative. We note that stability of the potential is ensured by some higher order operators and this is taken as a truncation of the full potential.
The scalar fields interact with gravity through the {the minimal gravitational coupling}
\begin{equation}
  {S} = \int \left( \frac{1}{16 \pi G} R + \mathcal{L} \right) \sqrt{-g} d^4x, 
\end{equation}
where $R$ is the Ricci scalar determined by the metric $g$ and $\mathcal{L}$ is given in Eq.~\ref{eq:Lagrangian-KG-curved-space}. Variation of the action with respect to the metric gives rise to Einstein equations
\begin{align}
    & R_{\mu \nu} - \frac{1}{2} R = 8 \pi G T_{\mu \nu},
\end{align}
where $R_{\mu \nu}$ is the Ricci tensor, and $T_{\mu\nu}$ the energy-momentum tensor given by 
\begin{align}
    T_\mu^\nu = & \sum_i \left (\frac{\delta \mathcal{L}}{\delta (\partial_\nu \phi_i)}\partial_\mu\phi_i + \frac{\delta \mathcal{L}}{\delta (\partial_\nu \phi_i^*)} \partial_\mu \phi_i^*
    \right ) - \delta^\nu_\mu \mathcal{L} \cr
     = &- \frac{1}{2} g^{\nu \nu'} \partial_{\nu'} \phi_1^* \partial_\mu \phi - \frac{1}{2} g^{\nu\nu'} \partial_{\nu'} \phi_1 \partial_\mu \phi_1^* - \frac{1}{2} g^{\nu \nu'} \partial_{\nu'} \phi_2^* \partial_\mu \phi_2 \cr
     &- \frac{1}{2} g^{\nu\nu'} \partial_{\nu'} \phi_2 \partial_\mu \phi_2^* + \delta_\mu^\nu \bigg ( \frac{1}{2} g^{\mu \nu}\partial_\mu \phi_1^* \partial_\nu \phi_1 +    \frac{1}{2} m_1^2 |\phi_1|^2  + \frac{1}{4}c_1|\phi_1|^4 \cr 
     &+\frac{1}{2} g^{\mu \nu}\partial_\mu \phi_2^* \partial_\nu \phi_2 +    \frac{1}{2} m_2^2 |\phi_2|^2 + \frac{1}{4}c_2|\phi_2|^4 + \frac{1}{4} c_{12} |\phi_1|^2|\phi_2|^2 \bigg ).
\end{align}
Note that since gauge fields are not the focus here, the simpler definition of $T^{\nu}_\mu$ is equivalent to the one using variation with respect to the metric. If we vary the action with respect to the scalar field we get the Klein-Gordon equation 
\begin{equation}
    g^{\mu \nu} \nabla_\mu \nabla_\nu \phi_i = \frac{d V}{d \phi_i^*},
  \end{equation}
  where $\nabla$ is the covariant derivative that contains the Christoffel symbols.

\subsection{Metric parametrization}
\label{sec:metr-param}

Assuming spherical symmetry of the metric,
we parametrize the metric as
\begin{align}
  \label{eq:metric}
  ds^2
  & =
    -B(r) dt^2 + A(r) dr^2 + r^2 d\theta^2 + r^2 \sin^2 \theta d
    \phi^2.
\end{align}
Solving the ${}^t_t$ and ${}^r_r$ components of the Einstein equation, we
have
\begin{align}
    \label{eq:KG-eom-1-2}
    & 
    \frac{4\pi G_N}{B(r)} \partial_t \phi_1 \partial_t \phi_1^* +  \frac{4\pi G_N}{A(r)} \partial_r \phi \partial_r \phi_1^* +\frac{4\pi G_N}{B(r)} \partial_t \phi_2 \partial_t \phi_2^* +  \frac{4\pi G_N}{A(r)} \partial_r \phi_2 \partial_r \phi_2^* \cr
    &\indent +4\pi G_N m_1^2 |\phi_1|^2   +4\pi G_N m_2^2 |\phi_2|^2 + 2\pi G_N c_1 |\phi_1|^4+ 2\pi G_N
    c_2 |\phi_2|^4+ 2\pi G_N c_{12}
    \phi_1|^2|\phi_2|^2 \cr 
    &\indent - \frac{A'(r)}{r A(r)^2} + \frac{1}{r^2 A(r)} - \frac{1}{r^2} = 0, \cr
    & \frac{4\pi G_N}{B(r)} \partial_t \phi_1 \partial_t \phi_1^* +  \frac{4\pi G_N}{A(r)} \partial_r \phi_1 \partial_r \phi_1^* + \frac{4\pi G_N}{B(r)} \partial_t \phi_2 \partial_t \phi_2^* +  \frac{4\pi G_N}{A(r)} \partial_r \phi_2 \partial_r \phi_2^* \cr
    &\indent -4\pi G_N m_1^2 |\phi_1|^2     -4\pi G_N m_2^2 |\phi_2|^2 - 2\pi G_N c_1 |\phi_1|^4 - 2\pi G_N c_2 |\phi_2|^4 - 2\pi G_N c_{12} |\phi_1|^2|\phi_2|^2 \cr 
    &\indent -\frac{B'(r)}{r A(r) B(r)} - \frac{1}{r^2 A(r)} + \frac{1}{r^2} = 0. \cr 
    & 
\end{align}
Two extra constraints come from the Klein-Gordon equations of motion.
Plugging in the covariant derivative, we get
\begin{align}
	\label{eq:KG-eom-3}
	&  \frac{1}{A}\partial_r^2 \phi_1 - \frac{1}{B}  \partial_t^2 \phi_1 +
	\partial_r \phi_1 \left (\frac{B'(r)}{2 A(r)B(r)} - \frac{A'(r)}{2A(r)^2} +
	\frac{2}{A(r)r}\right ) \cr 
	&\indent - m_1^2 \phi_1  - c_1 |\phi_1|^2 \phi_1 - \frac{1}{2} c_{12} |\phi_2|^2 \phi_1 = 0, \cr
	&  \frac{1}{A}\partial_r^2 \phi_2 - \frac{1}{B}  \partial_t^2 \phi_2 +
	\partial_r \phi_2 \left (\frac{B'(r)}{2 A(r)B(r)} - \frac{A'(r)}{2A(r)^2} +
	\frac{2}{A(r)r}\right ) \cr 
	&\indent - m_2^2 \phi_2 - c_2 |\phi_2|^2 \phi_2 - \frac{1}{2} c_{12} |\phi_1|^2 \phi_2 = 0.
\end{align}

\subsection{Rescaling to Dimensionless Variables}
\label{sec:resc-dimens-vari}
We take the harmonic ansatz with energy eigenstates $\phi_i(t,r) = \Phi_i(r)
\mathrm{e}^{-i\mu_i t}$. Plugging it into the equation of motion we can separate the
time evolution part of $\phi_i$. 
In order to solve it numerically, we perform the following rescaling of variables:
\begin{align}
      \Phi_1 & = \tilde \Phi_1 \; (4\pi G_N)^{-1/2} ,  & 
      \Phi_2 & = \tilde \Phi_2 \; (4\pi G_N)^{-1/2} , \cr
             \mu_1 & = \tilde \mu_1 \; m_1,  & 
             \mu_2 & = \tilde \mu_2 \; m_1, \cr
                      c_1 & = \tilde \lambda_1 \; 4 \pi G_N m_1^2, & 
                                c_2 & = \tilde \lambda_2 \; 4 \pi G_N m_1^2, \cr
                                          c_{12} & = \tilde
                                                         \lambda_{12}
                                                         \; 4 \pi G_N
                                                         m_1^2,  & 
                                                         m_{2}
  & = \tilde m_{r} m_1, \cr
r &  = \tilde r /m_1, & &    
\end{align}
where the variable with a tilde is dimensionless.\footnote{For comparison, in Ref.~\cite{Colpi:1986ye}, $\tilde r, \tilde \Phi, \tilde
\mu, \tilde \lambda$ are denoted as $x, \sigma, \Omega, \Lambda$
respectively.} If one parametrizes $c_i$ in the notation usually adapted in the axion literature using the Peccei-Quinn symmetry breaking scale, $c_i = m_i^2/f_i^2$, it reads
\begin{align}
  \tilde \lambda_1
  & =
    \frac{1}{4\pi} \left (\frac{M_{Pl}}{f_1} \right )^2, \cr
  \tilde \lambda_2
  & =
    \frac{1}{4\pi} \left (\frac{M_{Pl}}{f_2} \right )^2 \tilde m_r^2.
\end{align}
The strength of the interaction terms can be parameterized by the size of $f_i$. In other words, $\tilde \lambda \sim 1$ parametrizes a self-interaction whose strength is comparable to gravity with $f\sim M_{Pl}$. 
The dimensionless variables will be used for numerically solving the system. In what follows, we assume $m_1$, $m_2$ close to each other, so are $f_1$ and $f_2$. Therefore, we parametrize the coupling strength $c$'s in the unit of $m_1^2/f^2$ with $f$ chosen at $10^{17}\; \mathrm{GeV}$. In other words, we have
\begin{align}
  c_1
  & = \lambda_1 \frac{m_1^2}{f^2}, \cr
    c_2
  & = \lambda_2 \frac{m_1^2}{f^2}, \cr
    c_{12}
  & = \lambda_{12} \frac{m_1^2}{f^2}.
\end{align}
We will parametrize the three physical couplings with order one numbers in a tuple, $(\lambda_1, \lambda_2, \lambda_{12})$ in the rest of the paper. 
In terms of the dimensionless variables, we can write the equations of motion as
\begin{align}
  \label{eq:KG-eom-rescaled}
  & \left ( \frac{\tilde \mu_1^2}{B} + 1 \right )  \tilde
    \Phi_1^2 
    +  \frac{1}{A}  {\tilde \Phi_1^{\prime 2}}
    + \frac{1}{2}\tilde \lambda_1 \tilde
    \Phi_1^4
    +
    \left ( \frac{\tilde \mu_2^2}{B} + \tilde{m}_r^2  \right )  \tilde
    \Phi_2^2 
    +  \frac{1}{A}  {\tilde \Phi_2^{\prime 2}}
    + \frac{1}{2}\tilde \lambda_2 \tilde
    \Phi_2^4
    + \frac{1}{2}\tilde \lambda_{12} \tilde
    \Phi_1^2    \Phi_2^2 \cr
    &\indent - \frac{A'}{\tilde r  A^2} +
    \frac{1}{\tilde r^2 A} -
    \frac{1}{\tilde r^2} = 0, \cr
& \left ( \frac{\tilde \mu_1 ^2 }{B}  - 1 \right )\tilde \Phi_1^2
  +  \frac{1}{A} \tilde  \Phi_1^{\prime 2}
        - \frac{1}{2} \tilde \lambda_1 \tilde \Phi_1^4
+ \left ( \frac{\tilde \mu_2 ^2 }{B}  - \tilde{m}_r^2 \right )\tilde \Phi_2^2
  +  \frac{1}{A} \tilde  \Phi_2^{\prime 2}
        - \frac{1}{2} \tilde \lambda_2 \tilde \Phi_2^4
        - \frac{1}{2} \tilde \lambda_{12} \tilde \Phi_1 ^2 \Phi_2^2 \cr
&\indent -\frac{B'}{\tilde r A B} -
        \frac{1}{\tilde r^2 A} + \frac{1}{\tilde r^2}
        = 0, \cr
  & \frac{1}{A}{\tilde  \Phi_1^{\prime\prime }}
    +\left (\frac{\tilde \mu_1^2}{B} - 1 \right )\tilde \Phi_1 +
  \tilde \Phi'_1 \left (\frac{B'}{2 AB} - \frac{A'}{2A^2} +
  \frac{2}{A \tilde r}\right )  - \tilde \lambda_1 \tilde \Phi_1^3  -
    \frac{1}{2}    \tilde \lambda_{12} \tilde \Phi_2^2 \Phi_1 = 0,
        \cr
  & \frac{1}{A}{\tilde  \Phi_2^{\prime\prime }}
    +\left (\frac{\tilde \mu_2^2}{B}  - \tilde{m}_r^2 \right )\tilde \Phi_2 +
  \tilde \Phi'_2 \left (\frac{B'}{2 AB} - \frac{A'}{2A^2} +
  \frac{2}{A \tilde r}\right )  - \tilde \lambda_2 \tilde \Phi_2^3  -
    \frac{1}{2}    \tilde \lambda_{12} \tilde \Phi_1^2 \Phi_2 = 0.
\end{align}
 This concludes our setup of the problem, and we can solve Eq.~(\ref{eq:KG-eom-rescaled}) numerically. For details of the numerical algorithm, one can refer to Appendix~\ref{sec:NumericalProcedure}. Before we proceed to discuss the results and physical implications, we take a small detour to discuss the stability of solutions to Eq.~(\ref{eq:KG-eom-rescaled}).

\subsection{Time Evolution}
 
To verify that the solution is indeed stable against radial perturbations, we perform the time evolution using a finite difference method. This way, we can verify the temporal harmonic ansatz $\Phi_i(r,t) = \Phi_i(r) e^{-i\mu_it}$. The detailed procedure can be found in Appendix~\ref{sec:time-evolution}. We outline this procedure here briefly. 

\begin{figure}
  \centering
  \includegraphics[width=.45\textwidth]{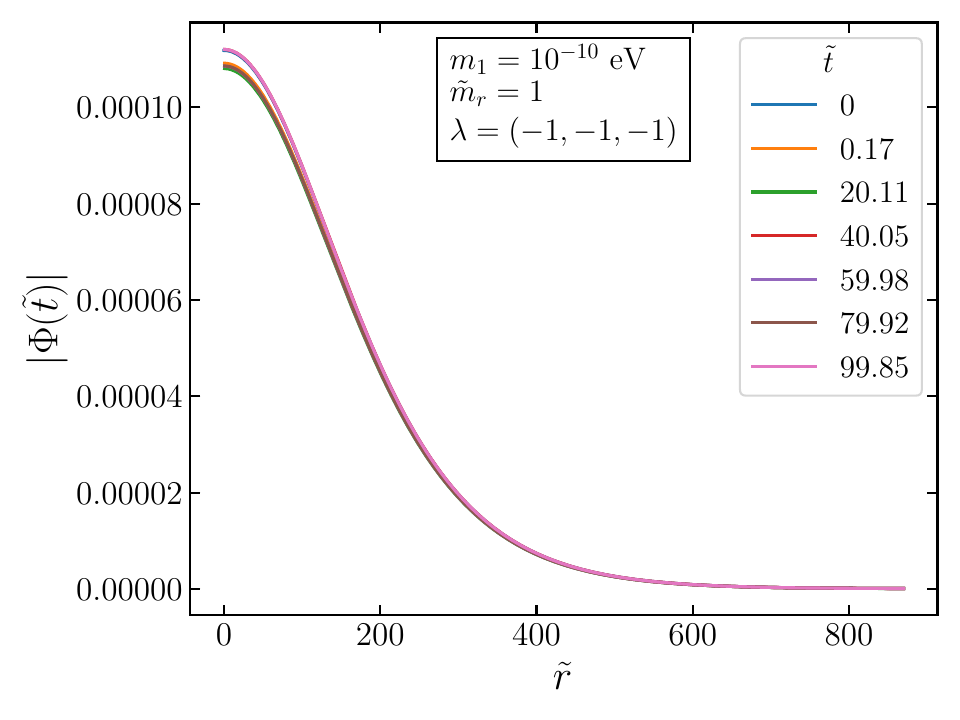}
    \includegraphics[width=.45\textwidth]{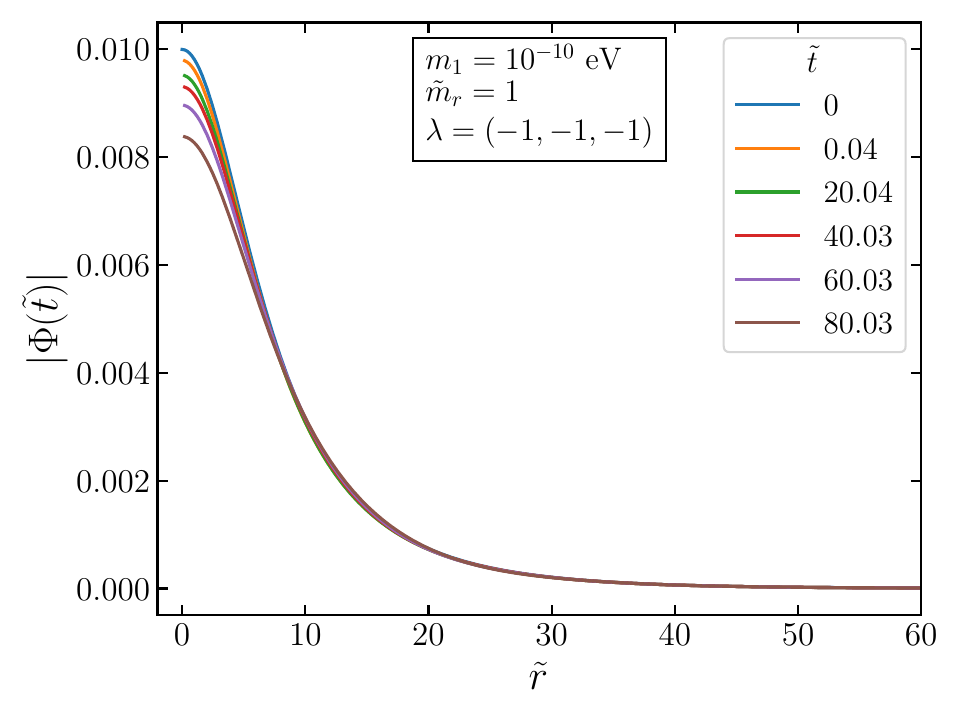}
    % \includegraphics[width=0.45\textwidth]{{Plots/time-evolve_1e-3}.pdf}
    % **** insert time evo for unstable one ****
    \includegraphics[width=0.45\textwidth]{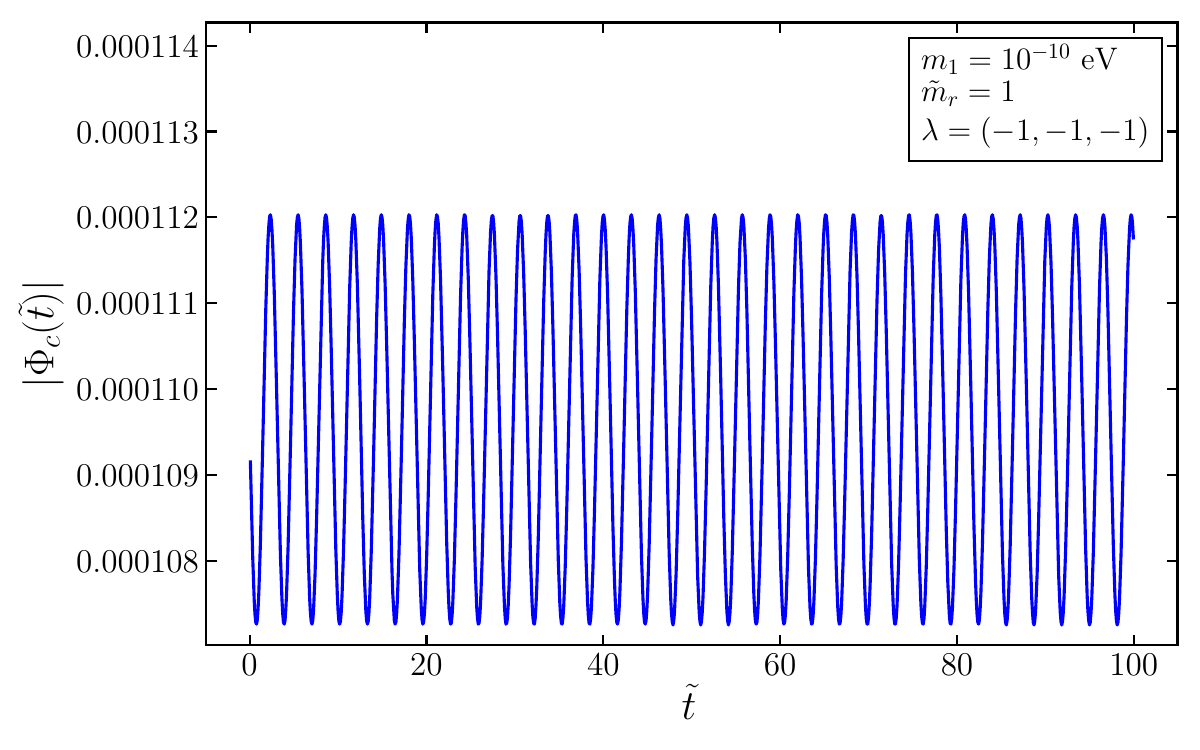}
    \includegraphics[width=0.45\textwidth]{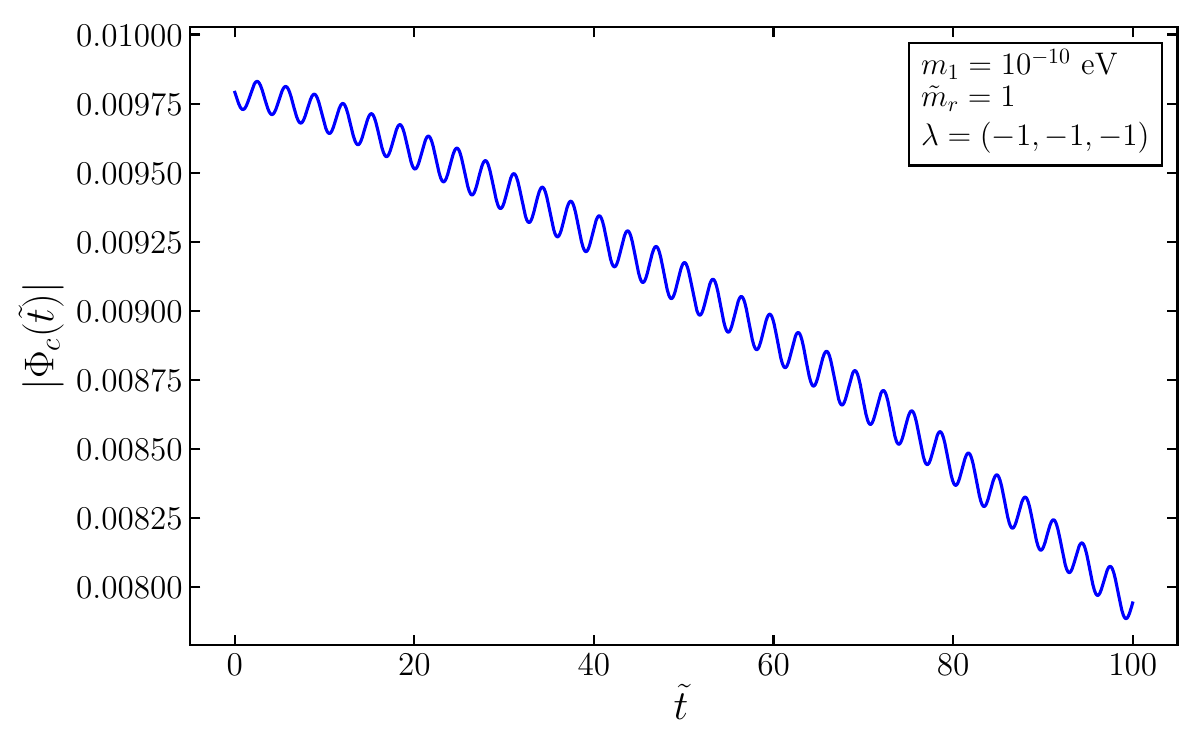}
    \caption{From top left clockwise: 1) time snapshots of the wave function of a stable configuration being radially perturbed $\psi \rightarrow \psi ( 1 + \epsilon ) $, with the vertical axis showing the sum of the modulus squared of the wave functions for $\Phi_1$ and $\Phi_2$ for $\lambda_1 = \lambda_2 = \lambda_{12} = -1$. 2) Same plot for an unstable configuration by increasing the central density controlled by $\Phi_1(0)$ and $\Phi_2(0)$. The wave function diverges quickly after a short time. 3) Unstable configuration in time domain shows the BEC collapsing within a short time period. 4) Stable configuration in time domain shows the system having small oscillation but maintaining a stable configuration. All the radial perturbations are done with $\epsilon \sim 2\%$. The stable configuration is chosen as $\tilde \Phi_{1}(0) = 10^{-4}$ and $\tilde \Phi_{2}(0) = 5\times 10^{-5}$; the unstable configuration $\tilde \Phi_{1}(0) = 10^{-4} $ and $\tilde \Phi_{2}(0) = 10^{-2}$.
    }
    \label{fig:time-evolve_1e-3}
  \end{figure}

First, one solves the static equations (\ref{eq:KG-eom-rescaled}). This will serve as the initial condition for the numerical time evolution. To check if it is a stable system, we perform a radial perturbation by $\Phi \rightarrow \Phi (1+\epsilon)$, and use $\Phi (1+\epsilon)$ as the initial condition instead. Here $\epsilon$ represents how far we perturb away from the static solution. In the stable case, the system can evolve for a long time with small oscillations, while in the unstable case the wave function quickly collapses or blows up depending on the sign of $\epsilon$. 
We show a sample stable solution in Fig.~\ref{fig:time-evolve_1e-3} for $\lambda_1 = \lambda_2 = \lambda_{12} = -1$ and initial central densities of $\tilde \Phi_{1}(0) = 10^{-4}$ and $\tilde \Phi_{2}(0) = 5\times 10^{-5}$, together with an unstable solution of $\tilde \Phi_{1}(0) = 10^{-4} $ and $\tilde \Phi_{2}(0) = 10^{-2}$.  % After a time of $\sim 40 m^{-1}$% ({\hl \textbf{CS:} I restored the unit of the time. Is this right, in the unit of $m^{-1}$?})
% , the initial central densities only deviate at most $10 \%$ from the static solution.
On top of the quasi-normal mode, there is no sign of decay after $\sim 100 m^{-1}$ in the stable scenario while the field with the unstable configuration decays considerably.

\subsection{Mass Profile of the BEC Structure}
With the solution to Eq.~(\ref{eq:KG-eom-rescaled}) in hand, one can derive the physical properties of the BEC system. The total mass of the BEC system is found by integrating the $T^{0}_0$ component of the energy momentum tensor
\begin{align}
	\label{eq:Mbs}
  M_{BS}
  &  \equiv 4 \pi \int_0^{\infty} dr r^2 T^0_0 \cr
  &  = 4 \pi \int_0^{\infty} dr r^2 
  \bigg( \frac{\mu_1^2}{2B} \Phi_{1}^2 + \frac{1}{2} m_1^2 \Phi_1^2  + \frac{1}{2A} \partial_r \Phi_1^2 + \frac{1}{4} \lambda_1 \Phi_1^4 + \cr
  & \qquad \qquad \qquad \qquad 
     \frac{\mu_2^2}{2B} \Phi_{2}^2 + \frac{1}{2} m_2^2 \Phi_2^2  + \frac{1}{2A} \partial_r
    \Phi_2^2 + \frac{1}{4} \lambda_2 \Phi_2^4
    + \frac{1}{4} \lambda_{12} \Phi_1^2 \Phi_2^2 \bigg).
\end{align}
In the linear regime, \textit{i.e.} non-relativistic limit, this reduces to the sum of the rest mass from the two scalars: $M_{BS} \approx M_1 + M_2 =  m_1 N_1 + m_2 N_2 + \mathcal{O}(B-1) + \mathcal{O}(1-A)$, with the subscript $BS$ standing for BEC system. 
 The compactness of the BEC system is defined as the mass-radius ratio in natural units
\begin{equation}
    C_{BS} = \frac{G_N M_{BS}}{R_{90}}
\end{equation}
where $R_{90}$ is the radius that contains $90\%$ of the total mass. Similar to the mass components, in the limit of small interaction between $\Phi_1$ and $\Phi_2$, we can also define $C_{1}$ and $C_{2}$ as the compactness of the BEC component corresponding to $\Phi_1$ and $\Phi_2$, respectively.  For a given set of model parameters, $(\lambda, m)$, in the single scalar scenario, the mass and compactness are solely determined by a single variable, $\Phi(r=0)$, which is related to the central density of the BEC system. As a result, the compactness is a function of mass. In the case of two scalar BEC, this is no longer true. One can find solutions with different combinations of $\Phi_1(0), \Phi_2(0)$, which essentially enlarges the parameter space to being 2-dimensional. The mass profile in the $C_{BS}-M_{BS}$ plane is no longer a curve, but a 2D region. A detailed analysis for various BEC mass profiles will be given in the results section of Section~\ref{sec:galactic-scale-bec} and \ref{sec:BosonStars} respectively. 

\subsection{Non-Relativistic Limit}
\label{sec:non-relativistic}
Although we solve the system in the full relativistic regime, expanding in the weak gravity limit can bring insights to various properties of the system. The waveforms of $\Phi_{1}$ and $\Phi_2$ should decay to zero as $r \rightarrow \infty$. % This means that the waveforms should decay exponentially for large r with terms of higher order in $\Phi^2$ negligible.
We also expect that the metric becomes non-relativistic for large $r$ such that
\begin{align}
  A(r)
  &= 1 - 2 V(r)\cr 
    B(r)
  &= 1+ 2V(r) \cr
    V(r)
  &= - \frac{G M(r)}{r} = -\frac{G (M_1(r) + M_2(r))}{r}.
\end{align}
We adopt the following ansatz for the wave function, which is verified by the numerical computation to hold well enough at sufficiently large radius. 
\begin{equation}
	\Phi_i = \sqrt{\frac{N_i}{\pi m_i R_i^3}} e^{- r/R_i}
\end{equation}	
where $N_i$ is the total number of particles for either $\Phi_{1}$ or $\Phi_2$ and $R_i$ is characteristic size of each BEC clump. Using this non-relativistic ansatz, we may integrate each term individually in Eq.~\ref{eq:Mbs} and combine terms to give the non-relativistic Hamiltonian.
In the weak gravity limit, one can parametrize the eigenvalues as
\begin{align}
  \mu_i
  & = m_i \left ( 1-  \alpha V(r) \right ).
\end{align}
Matching to the non-relativistic results in the leading terms, we found $\alpha=5/4$ reproduces the self-gravity term as in the single scalar case \cite{Croon:2018ybs}. We can then write out the Hamiltonian as
\begin{align}
  \label{eq:Hnr}
  H_{kin}
  =
  &
    +\frac{N_1}{2 m_1 R_1^2}
    +\frac{N_2}{2 m_2 R_2^2}
    \cr
    H_{int}
    =
  &
    +\frac{\lambda _1 N_1^2}{32 \pi  f^2 R_1^3}
    +\frac{\lambda _2 N_2^2}{32 \pi  f^2 R_2^3}
    +\frac{\lambda _{12} N_1 N_2}{4 \pi  f^2 \left(R_1+R_2\right){}^3} 
    \cr
    H_{grav}
    =
  &
    -\frac{5 Gm_1^2 N_1^2}{16 R_1}
    -\frac{5 G m_2^2 N_2^2}{16 R_2}
    -\frac{Gm_1m_2N_1N_2}{R_{\rm eff}}
    \cr
    H_{metric}
    =
  &
    - \frac{5 G N_1^2}{16 R_1^3}
    - \frac{5 G N_2^2}{16 R_2^3}
    - \frac{GN_1N_2}{\tilde R_{\rm eff}^3}
\end{align}
with
\begin{align*}
  \frac{1}{R_{\rm eff}}
  & =
    \frac{5}{8R_1} + \frac{5}{8R_2}
    - \frac{1}{R_1 + R_2}
    - \frac{R_1 R_2}{(R_1 + R_2)^3},
    \cr
    \frac{1}{\tilde R_{\rm eff}^3}
  & =
    \frac{m_1(4R_1 + R_2)}{m_2(R_1+R_2)^4}
    + \frac{m_2(R_1 + 4R_2)}{m_1(R_1+R_2)^4}.
\end{align*}
% \begin{align}
%   \label{eq:Hnr}
%   H_{kin}
%   =
%   &
%     +\frac{N_1}{2 m_1 R_1^2}
%     +\frac{N_2}{2 m_2 R_2^2}
%     \cr
%     H_{int}
%     =
%   &
%     +\frac{\lambda _1 N_1^2}{32 \pi  f^2 R_1^3}
%     +\frac{\lambda _2 N_2^2}{32 \pi  f^2 R_2^3}
%     +\frac{\lambda _{12} N_1 N_2}{4 \pi  f^2 \left(R_1+R_2\right){}^3} 
%     \cr
%     H_{grav}
%     =
%   &
%     -\frac{5 Gm_1^2 N_1^2}{16 R_1}
%     -\frac{5 G m_2^2 N_2^2}{16 R_2} 
%     \cr
%     H_{metric}
%     =
%   &
%     - \frac{5 G N_1^2}{16 R_1^3}
%     - \frac{5 G N_2^2}{16 R_2^3}
%     - \frac{5 Gm_1 m_2 N_1 N_2 R_1^4}{8 R_2(R_1+R_2)^4}
%     - \frac{5 Gm_1 m_2 N_1 N_2 R_2^4}{8 R_1(R_1+R_2)^4}
%     - \frac{17 Gm_1 m_2 N_1 N_2 R_1^3}{8(R_1+R_2)^4}      
%     \cr
%   &
%     - \frac{17 Gm_1 m_2 N_1 N_2 R_2^3}{8(R_1+R_2)^4}
%     - \frac{9 Gm_1 m_2 N_1 N_2 R_1^2R_2}{4(R_1+R_2)^4}
%     - \frac{9 Gm_1 m_2 N_1 N_2 R_1R_2^2}{4(R_1+R_2)^4}
%     \cr
%   &
%     - \frac{Gm_2 N_1 N_2 R_1}{m_1(R_1+R_2)^4}
%     - \frac{Gm_1 N_1 N_2 R_2}{m_2(R_1+R_2)^4}
%     - \frac{4Gm_1 N_1 N_2 R_1}{m_2(R_1+R_2)^4}
%     - \frac{4Gm_2 N_1 N_2 R_2}{m_1(R_1+R_2)^4}.        
% \end{align}
We can immediately recognize the familiar form of kinetic terms, self-interaction terms, self-gravity terms, and the first two terms in $H_{metric}$, which are due to the kinetic term in the curved space-time consistent with the result in \cite{Croon:2018ybs,Guo:2019sns}. However, the Hamiltonian also contains a term proportional to $\lambda_{12} N_1 N_2$ that is due to the non-gravitational interaction between the two scalars. In addition, there are terms proportional to $G_NM_1 M_2$ due to the gravitational interaction between the two scalars which are not present in the single scalar case.  One recovers the one scalar result if either $N_1$ or $N_2$ is set to zero. In the limit of $\lambda_{12} \rightarrow 0$ and $R_1, R_2 \sim R$, $1/R_{\rm eff}$ reduces to $(5/8R)$, which is consistent with the result of \cite{Eby:2020eas} up to the $H_{grav}$ term.

\begin{figure}[thb]
    \centering
    \includegraphics[width=0.6\textwidth]{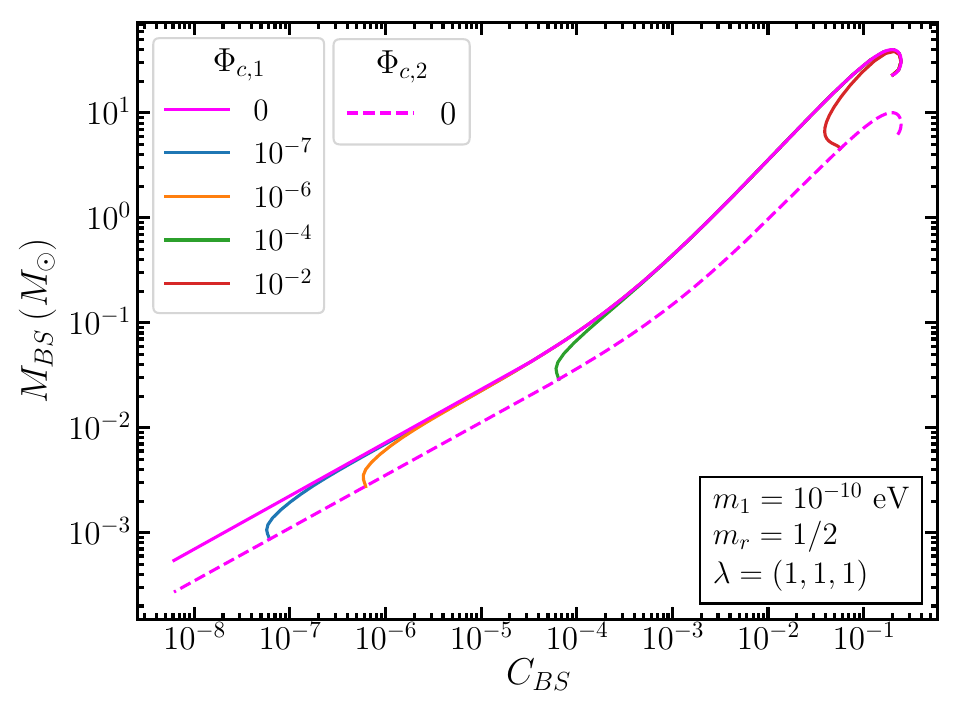}
    \caption{The total mass vs compactness for various values of $\Phi_{1}(0)$ and $\Phi_{2}(0)$ with $\lambda_1 = 1$, $\lambda_2 = 1$, and $\lambda_{12} = 1$. {\hl We scan $\Phi_{1}(0)$ and $\Phi_{2}(0)$ to show the existence of stable solutions. }
The solid magenta line corresponds to the single scalar limit with $m = 5 \times 10^{-11}$ eV and the dashed magenta line is the single scalar limit with $m = 10^{-10}$ eV. The solid lines in between corresponds to $m_1 = 10^{-10}$ eV and $m_2 = 5\times 10^{-11}$ with each curve corresponding to a different fixed $\Phi_{1}(0)$ and scanning over $\Phi_{2}(0)$. We let $\Phi_1$ dominate first, and $\Phi_2$ starts to dominate the system at different places due to $\Phi_1(0)$ fixed at different values in blue, orange, green, and red curves. One can see that the moment $\Phi_2$ starts to dominate, the curve transition from the dashed to solid magenta as expected. 
    }
    \label{fig:cvm-color}
\end{figure}

A discussion based on the Hamiltonian is in order. In the linear regime, one can safely neglect the terms in $H_{int}$ and $H_{metric}$. 
Without loss of generality, when $N_1 \gg N_2$, the gradient term will balance the gravity term Eq.~\ref{eq:Hnr} to give 
\begin{equation}
  \label{eq:linear-scaling}
    N \approx N_1 \propto \frac{M_{Pl}^2}{m_1^3 R_1}.
\end{equation}
Whether $\Phi_1$ or $\Phi_2$ dominates is solely determined by the central value, $\Phi_1(0)$ and $\Phi_2(0)$. Therefore, we have
\begin{align}
  M_{BS} \sim
  \begin{cases}   
    \frac{M_{Pl}^2}{m_1} \sqrt{C}, & N_1 \gg N_2 \\
    \frac{M_{Pl}^2}{m_2} \sqrt{C}, & N_2 \gg N_1 
  \end{cases}
\end{align}
It is expected that when one goes from the  $\Phi_1$-dominating regime to $\Phi_2$-dominating regime, the $M-C$ relation will smoothly transition from one to the other. This is verified by the numerical computation and shown in Fig.~\ref{fig:cvm-color}. This has interesting implications for the solitonic core, in the context of galactic scale BEC. We will discuss this feature in more detail in Section~\ref{sec:galactic-scale-bec}. Note that we only show the result with $m_2/m_1 = 1/2$ due to the numerical complexity, but in principle the result should hold for much larger scalar ratios. This means that by adjusting the central densities of the two BEC components, $\Phi_1(0)$ and $\Phi_2(0)$, one can have non-trivial mass profiles that cannot be mimicked by a single scalar BEC system.

\begin{figure}[t]
  \centering
  \includegraphics[width=.45\textwidth]{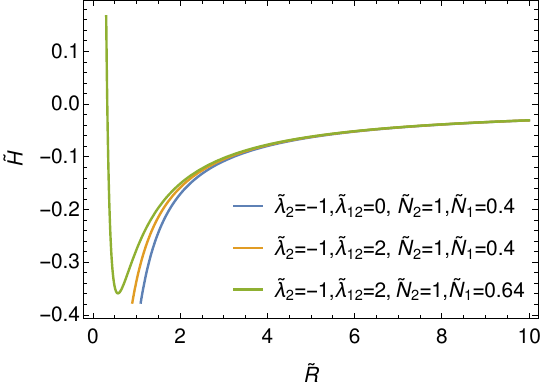}
  \includegraphics[width=0.45\linewidth]{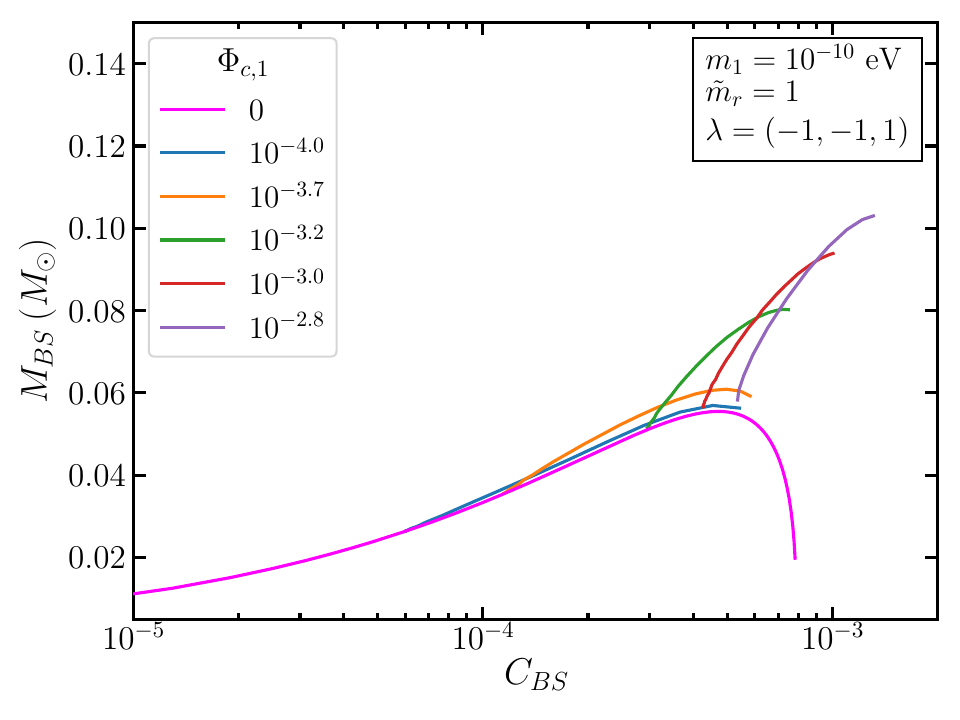}  
  \caption{Left: the rescaled Hamiltonian as a function of the BEC radius $\tilde R$ in the nonlinear regime, and $N_2 > N_1$. In particular, we observe that when there is no interaction between $\Phi_1$ and $\Phi_2$, it behaves the same as the single scalar case where $-\Phi_2^4$ destroys the local minimum so the system is not stable (blue curve). When (a) there is a repulsive interaction $\lambda_{12}>0$, and (b) the sub-dominant scalar number, $N_1$, is large enough, one can see the local minimum is restored (green curve). This happens only when both conditions are fulfilled. With only (a), the system still lacks local minimum (orange curve). Right: The numerical solutions verifies the previous analysis. As an example, we choose $\tilde \lambda_1 = 1, \tilde \lambda_2 = -1, \tilde \lambda_{12} = 1$. All curves are generated by scanning over $\Phi_2(0)$. The magenta curve is with only $\Phi_2$ field. One can see the there are no more solutions beyond $C_{BS}\gtrsim 8\times 10^{-4}$ due to $-\Phi_2^4$ self-interaction. In the colored curves we add a sub-component $\Phi_1$ with $\Phi_1(0)$ fixed to the labeled value, and again scan over $\Phi_2(0)$. When the subdominant scalar $\Phi_1$ number is large enough (green, red, purple), the system can becomes stable again and one can find solutions at $C_{BS}>10^{-3}$, consistent with our analytical approximation. {\hl We scan $\Phi_{2}(0)$ to show the existence of stable solutions. }}
  \label{fig:HR-nonlinear}
\end{figure}

In the nonlinear regime, the non-gravitational interactions are important. In contrast to the single scalar case, we have three non-gravitational interaction terms proportional to $\lambda_1, \lambda_2,$ and $\lambda_{12}$ respectively. It is known that a repulsive self-interaction $+\Phi^4$ stabilizes the system \cite{Colpi:1986ye,PhysRevD.38.2376,Schiappacasse:2017ham,Croon:2018ybs,Guo:2019sns}, while the attractive $-\Phi^4$ renders the system unstable once it goes to the nonlinear regime \cite{Eby:2016cnq,Schiappacasse:2017ham,Croon:2018ybs}. In the two scalar system, we observe that a repulsive non-gravitational interaction $+\Phi_1^2 \Phi_2^2$ also stabilizes the system. This can be understood by looking into the Hamiltonian of the system. In the nonlinear regime, the important terms are the gravity potential and the nonlinear terms. We assume $\Phi_2$ dominates the system, $N_2 > N_1$:
\begin{align}
  \label{eq:1}
  \tilde H(R_2)
  & \approx
    \frac{\tilde \lambda_2 \tilde N_2^2}{8 \tilde R_2^3} +   \frac{\tilde \lambda_{12} \tilde N_2 \tilde N_1}{8 \tilde R_2^3}
    - \frac{5 \tilde N_2^2}{16 \tilde R_2},
\end{align}
where $H = \tilde H m\Delta^2$, $N = \tilde N (m/M_{Pl})^2  \Delta$, with $\Delta$ being some large number for normalization.
{\hl If we start with $\tilde \lambda_1 < 0, \tilde \lambda_2< 0, \tilde \lambda_{12} > 0$ ($\tilde \lambda_1 > 0, \tilde \lambda_2 > 0, \tilde \lambda_{12} < 0$), we observe that the BEC system can be stabilized (destabilized), respectively, when the following condition is met:
\begin{align}
\frac{\tilde N_1 } { \tilde N_2} >  \left | \frac{\tilde \lambda_2 }{\tilde \lambda_{12}} \right |.
\end{align}
We demonstrate this both analytically and numerically in Fig.~\ref{fig:HR-nonlinear}. This has some interesting implications for boson stars. 
Contrary to the common understanding that BECs resulting from $-\phi^4$ or $\Lambda^4 (1-\cos(\phi))$ potentials (\textit{e.g.} axion stars) are dilute, they can be stabilized up to high density if there are multiple of them and different species interact with each other through a repulsive interaction.
}
% , even if we start with attractive self-interactions for both scalars $\tilde \lambda_1, \tilde\lambda_2<0$.
% We stress that this makes model building easier as $-\Phi^4$ interactions are common in axion models and it is shown to be nontrivial to realize $+\Phi^4$ type of interactions \cite{Fan:2016rda}.

% {\hl
%   We stress that at 
%   }

In other combinations of $\lambda$'s, the presence of the coupling between the two scalars can also offer unique features in the $C_{BS}-M_{BS}$ curves in the non-linear regime that are not possible in the single scalar limit.
This has interesting implications for the stellar scale BEC structure. We will discuss this application in more detail in Section~\ref{sec:BosonStars} and show the results for how the nonlinear regime is changed by varying the model parameters, $\lambda_{1}$, $\lambda_2$, $\lambda_{12}$, and $m_1$,$m_2$.  
\section{Galactic Scale BEC Structure}
\label{sec:galactic-scale-bec}
Having discussed the behavior of the two scalar BEC system, we now turn to applications: the first being the implications at galactic scales. We first briefly review the problem of using single scalar BEC to fit  galaxy data. For details one can refer to e.g. \cite{Deng:2018jjz}. We then show that with the second scalar and the transitioning behavior demonstrated in Section~\ref{sec:non-relativistic}, both the best fit and the data points themselves can be accommodated in the two scalar scenario.

\subsection{The Problem with a Single Scalar BEC}
\label{sec:problem-with-single}
While the NFW profile describes dark matter halo density to good precision at radii larger than $\sim \mathrm{kpc}$, it is known that at sub-kpc distances the densities of typical galaxies approach constant values. Measurements of galaxy rotation curves yield the profile of the core density and core size across galaxies of different sizes \cite{Donato:2009ab,Salucci:2018hqu,Rodrigues:2017vto,Deng:2018jjz}. One can parametrize the density by
\begin{align}
  \rho(r)
  & =
    \frac{\rho_c}{1+r^2/R^2_c},
\end{align}
where $R_c$ is the radius where the density drops to half of the central density. By comparing with the measurement in \cite{Rodrigues:2017vto}, one can see the core size and core density can be fitted by
$\rho_c \propto 1/R_c^\beta$, with $\beta \approx 1.3$ \cite{Deng:2018jjz}. 
On the other hand, in a single scalar system, the scaling behavior is completely fixed by the scalar potential {\hl and dynamics can not change the mass-radius relation.} For example, in the linear regime, it is $H_{kin}$ balancing with $H_{grav}$, which gives the scaling shown in Eq.~(\ref{eq:linear-scaling}). This translates to $\rho \sim 1/R^4$. Similarly, one can try different polynomial potentials, and they result in different $\beta$'s, but none of them {\hl falls into $1 \lesssim \beta \lesssim 1.3$}  even if one takes into account the scattering of the data. The values of the scaling index $\beta$ are summarized in Table~\ref{tab:scaling-index}. This poses a challenge to using ultralight dark matter to address the core-cusp problem \cite{Deng:2018jjz}, which was one of the main motivations of ultralight dark matter \cite{Hu:2000ke,Hui:2016ltb}. 
\begin{table}[th]
  \centering
\begin{align*}
  \begin{array}[ht]{|c|c|c|c|}
    \hline
    & H_{kin} & \phi^4 & \phi^6 \\
    \hline
    H_{grav} &  \beta = 4 & \beta = \infty & \beta = -2\\
    \hline
    -\phi^4 & (\beta = 2)& \text{n/a} & \beta = 0 \\
    \hline
\end{array}
\end{align*}
\caption{The scaling index $\beta$ for one scalar BEC with different scalar potentials. In any given regime, the system is balanced by two dominating terms, one in the top row and one from the left column. Note that $-\phi^4$ balancing $H_{kin}$ is not stable.}
  \label{tab:scaling-index}
\end{table}
{\hl
  As we have briefly discussed, this problem can be understood as follows. When the system is composed of one scalar, the Hamiltonian can be written as
  \begin{align}
      H
    \approx
    &
    \;\frac{N_1}{2 m_1 R_1^2}     +\frac{\lambda _1 N_1^2}{32 \pi  f^2 R_1^3}      -\frac{5 Gm_1^2 N_1^2}{16 R_1}     - \frac{5 G N_1^2}{16 R_1^3},
  \end{align}
  which leads to a fixed $M(R)$ relation. There is no room for the scalar dynamics to alter this relation, as both mass and radius are parametrized by the $\phi(0)$. 
}

\subsection{Two-Scalar BEC to the Rescue}
\label{sec:two-scalar-bec}
{\hl In this section, we first allow $\Phi_1(0)$ and $\Phi_2(0)$ to vary freely, and show that, while maintaining radial stability, there is parameter space that can accommodate the galactic data. We then comment on the implication for the dynamics of the two scalars.}

\begin{figure}[t]
  \centering
  \includegraphics[width=.32\textwidth]{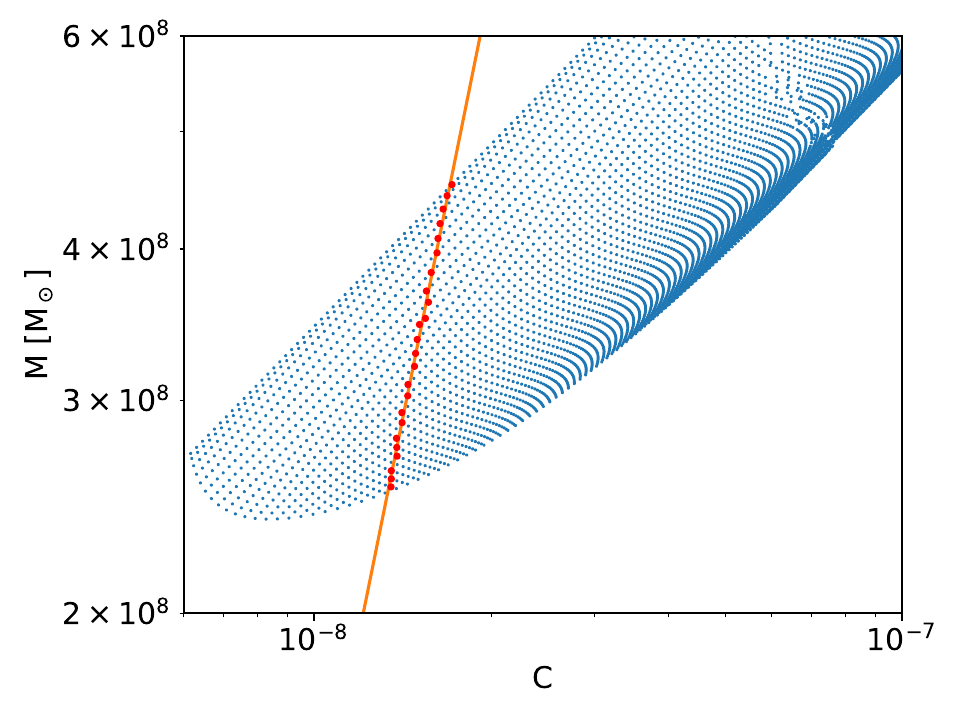}
  \includegraphics[width=.32\textwidth]{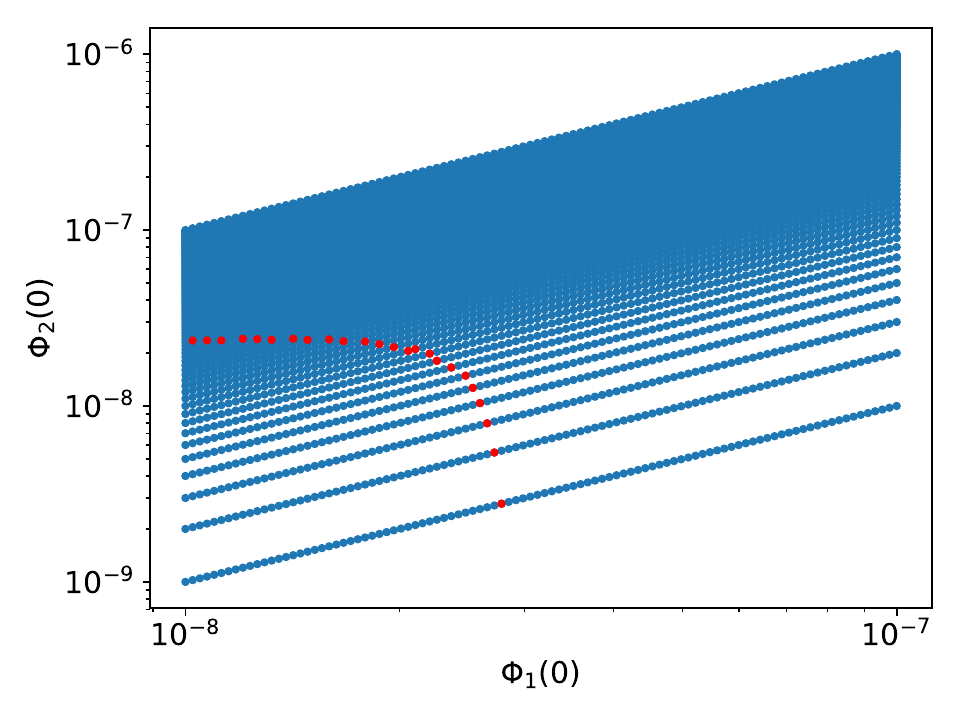}
  \includegraphics[width=.32\textwidth]{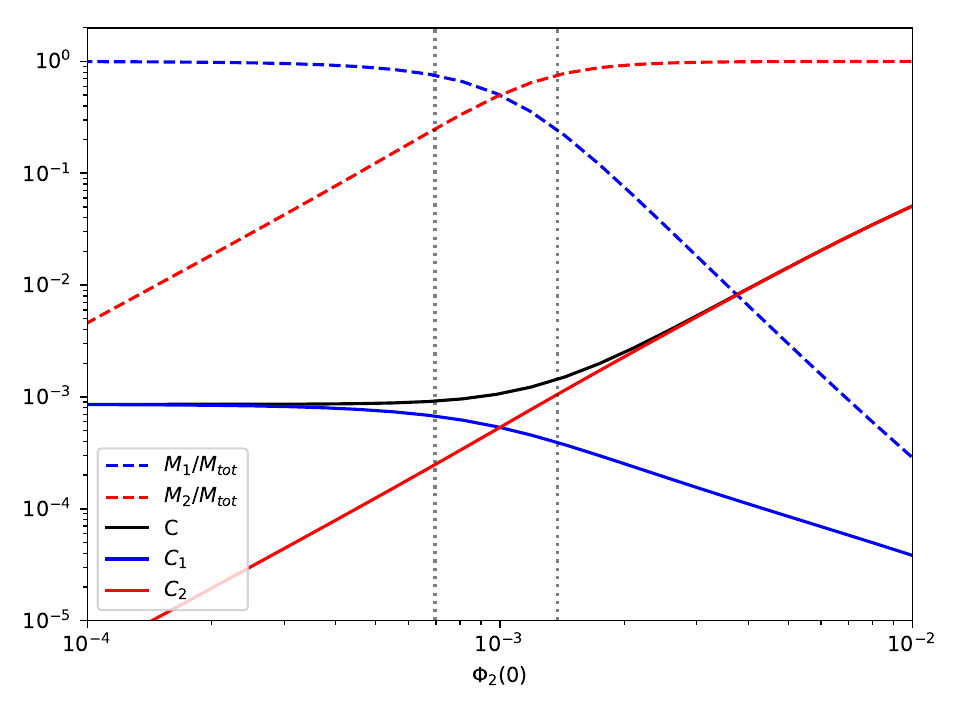}  
  \caption{Left and middle: this is one example numerical scan over $\Phi_1(0)$, $\Phi_2(0)$ (blue points). Each blue point in the middle plot gives a unique mass profile, \textit{i.e.} a blue point in the left panel in the $C-M$ space. The orange line is the best fit taken from \cite{Deng:2018jjz}. From the numerical data cloud we pick the points to reconstruct the orange curve. These are shown as the red points in both panels. This particular scan is conducted with $m_1=10^{-22} \;\mathrm{eV}, m_2=2\times 10^{-22} \;\mathrm{eV}$, and $\lambda_1 = \lambda_2 = \lambda_{12}=0$. {\hl We scan $\Phi_{1}(0)$ and $\Phi_{2}(0)$ to show the existence of stable solutions.  Right: we show the transition region from $\Phi_1$ dominating to $\Phi_2$ dominating by taking one slice of the scan with fixed $\Phi_1(0)$. One can see that indeed when one component mass is more than $75\%$ the total mass, the compactness of the system is mostly determined by this mass component. We define the region between the grey dotted region to be the transition region.}}
  \label{fig:beta-scan}
\end{figure}

We note that in the case of a two scalar system with $m_1 \neq m_2$, the curve in the $C-M$ plane is a smooth interpolation of the mass profile of each scalar as shown in Fig.~\ref{fig:cvm-color}. Scanning over the $\Phi_1(0)- \Phi_2(0)$ space gives us a region in the $C-M$ plane, with one-to-one correspondence of each $(\Phi_1(0), \Phi_2(0))$ point to a $(C, M)$ point.
On the other hand, galactic data points can be fit with a curve of $M \sim C^{2.4}$, \textit{i.e.} $\rho \sim 1/R^{1.3}$. Therefore, by looking at the $(\Phi_1(0), \Phi_2(0))$ to $(C, M)$ correspondence,  one can find the one-dimensional curve that reproduces the best fit of the data. We show this explicitly in Fig.~\ref{fig:beta-scan}. We observe that a curve in $\Phi_1(0), \Phi_2(0)$ space can {\hl accommodate} the best fit. {\hl In particular, we observe the mass profile of the total BEC is mostly dominated by one component if the component weighs more than $75\%$ percent of the total mass. This indeed is a useful criteria in determining the dominant BEC component, as starting from this point the compactness is mainly determined by the dominant species. We show this in the right plot of Fig.~\ref{fig:beta-scan}.}

% \begin{figure}[h]
%   \centering

%   \caption{In the plot we }
%   \label{fig:M1-M2}
% \end{figure}

\begin{figure}[t]
  \centering
  \includegraphics[width=.45\textwidth]{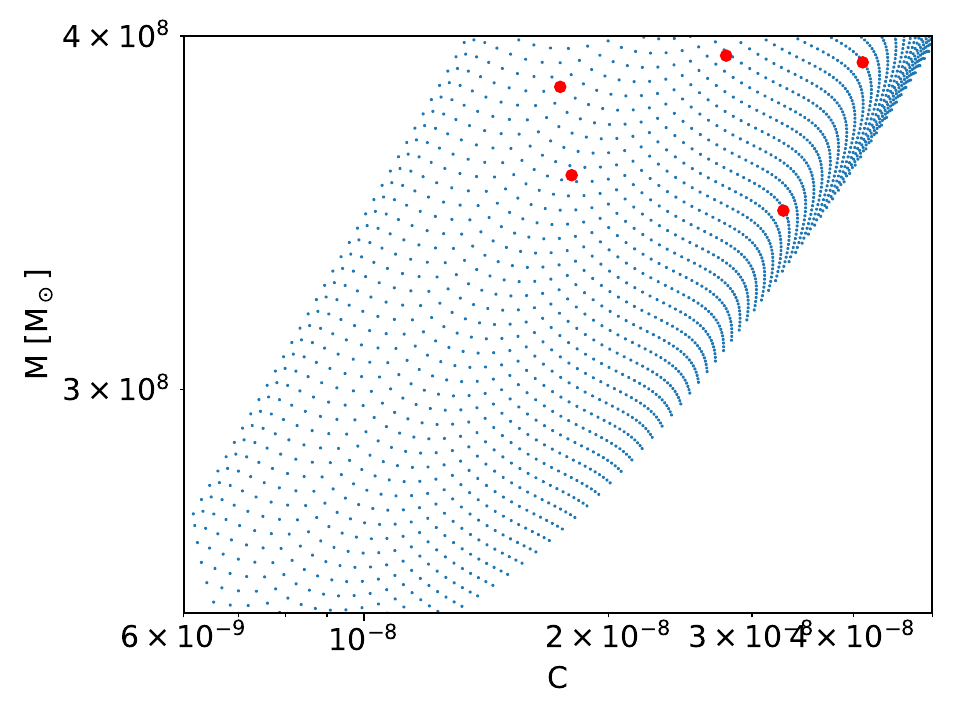}
  \includegraphics[width=.45\textwidth]{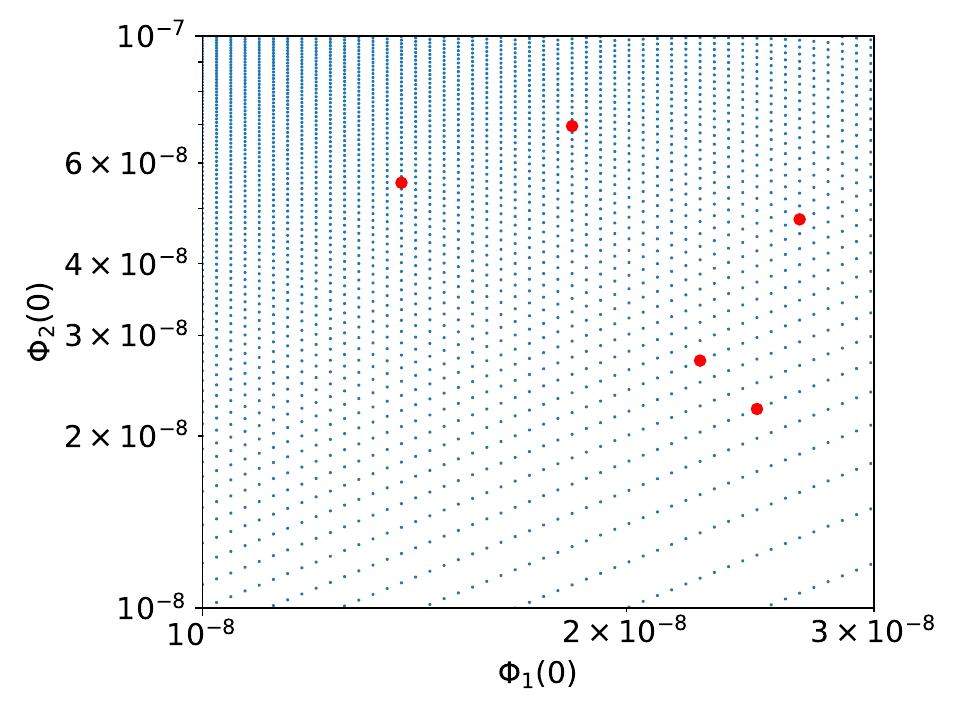}
  \caption{We show the numerical scan (blue) in both $C-M$ space (left) and $\Phi_1(0), \Phi_2(0)$ space (right). The red points are a few data points taken from \cite{Rodrigues:2017vto}. This particular scan is conducted with $m_1=10^{-22} \;\mathrm{eV}, m_2=2\times 10^{-22} \;\mathrm{eV}$, and $\lambda_1 = \lambda_2 = \lambda_{12}=0$. See the main text for more details. {\hl We scan $\Phi_{1}(0)$ and $\Phi_{2}(0)$ to show solutions exist while the system maintains radial stability. }}
  \label{fig:beta-scan-datapoints}
\end{figure}

As we discussed in Section~\ref{sec:problem-with-single}, with two scalars one can {\hl accommodate the galactic data points even after taking into account the scattering of the data set,} instead of just the $\rho \sim 1/R^{1.3}$ curve that best fits the data points. This can be done with the following procedure: we overlay the galactic data points with the numerical scan in the $C-M$ space (or equivalently in the $\rho-R$ space.) Then for each data point we can find a numerical point that is identical up to the scan resolution. We can then go back to the $\Phi_1(0), \Phi_2(0)$ space and identify the value for the  $\Phi_1(0), \Phi_2(0)$ pair needed to reproduce this data point. The result is shown in Fig.~\ref{fig:beta-scan-datapoints}.

{\hl We emphasize that, however, this does not mean the two scalar BEC can accommodate any galactic data, or lacks predictability. While our approach only considers radial stability and existence of solutions of the BEC system, this exercise translates the galactic data to a requirement of the field value in the classical field configuration space. This requirement, shown in the right panels of Figs.~(\ref{fig:beta-scan}-\ref{fig:beta-scan-datapoints}), can be further constrained after the two scalar dynamics is taken into account. In addition, because of the extra scalar, there could be new constraints specific to the two scalar system, such as that related to the relaxation time scale. Since in this work we focus on the radial stability of the BEC system, we leave the study dedicated for constraining the two scalar BEC with galactic data for the future.
}

\begin{figure}[t]
  \centering
  \includegraphics[width=.45\textwidth]{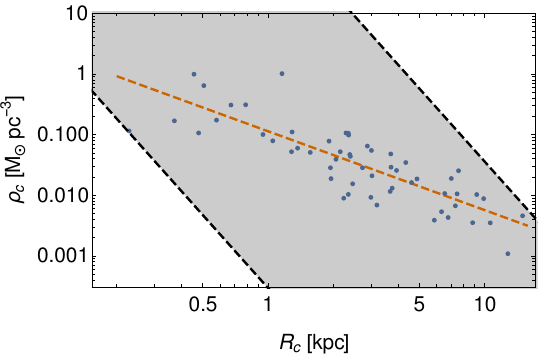}
  \caption{% Left: numerical scan of $\Phi_1(0)$ and $\Phi_2(0)$. The scan shows that in the two scalar scenario it is possible to fill the gap between the two single scalar limits with certain combinations of $\Phi_1(0)$ and $\Phi_2(0)$.
We extrapolate the numerical results to a larger mass separation, $m_1 = 2\times 10^{-24}\;\mathrm{eV}$ and $m_2 = 2\times 10^{-21}\; \mathrm{eV}$. The grey region is what the two scalar BEC system covers with varying central density determined by $\Phi_1(0)$, $\Phi_2(0)$. One can see that it not only contains the best fit curve (orange) that represents $\rho_c \approx 1/R_c^{1.3}$, it also incorporates the scattering of the data. The data points are from \cite{Rodrigues:2017vto}, and fit adopted from \cite{Deng:2018jjz}. {\hl We extrapolate the scalar mass gap $m_1/m_2$, while use a range in the BEC fraction, $M_1/M_2$, comparable to our numerical computations with smaller $m_1/m_2$ to show the existence of stable solutions. }}
  \label{fig:scan-of-phi1-phi2}
\end{figure}

Lastly, we can extrapolate the numerical scan to the scenario where two scalars have a larger mass ratio. Given that we know the BEC system behaves like a single scalar when $\Phi_1(0)\gg \Phi_2(0)$ or vice versa, we can estimate how big a mass ratio is needed to accommodate all the data points. The extrapolation result is shown in Fig.~\ref{fig:scan-of-phi1-phi2} with $m_1 = 2\times 10^{-24}\;\mathrm{eV}$ and $m_2 = 2\times 10^{-21}\; \mathrm{eV}$. 
{\hl This can be further used to constrain the two scalar model.} At its face value, it seems that Lyman-alpha \cite{Irsic:2017yje,Nori:2018pka} and subhalo mass function\cite{Schutz:2020jox} might heavily constrain the lighter of the two scalar. However, it is known that the nonlinear interaction terms could play an important role in the evolution of cosmic perturbation and structure formation \cite{Arvanitaki:2009fg,Arvanitaki:2019rax,Fan:2016rda}. %A thorough study of the suppression of matter power spectrum is beyond the scope of this work. %We hope to return to this in the future.
{\hl % Some comments on the mass ratio of the two BEC components are in order.
  The extrapolation shown in Fig.~\ref{fig:scan-of-phi1-phi2} is in $m_1/m_2 \sim \mathcal{O}(10^{-3})$. If one fixes the radius, the BEC ratio $M_1/M_2$ needs to change three orders of magnitudes to go from $\phi_1$ dominating the system to $\phi_2$ dominating the system. More specifically, one needs that at $R_c \sim 10 \, \mathrm{kpc}$, scalar one dominates, and at $R_c \sim 0.5\, \mathrm{kpc}$ scalar two dominates.
  % The mass ratio of the two components needed to dominate the system, if their behavior follows what we observed numerically with a smaller $m_1/m_2$, only needs to vary $\mathcal{O}(1)$. In other words, one can safely assume the system to be dominated by the bigger component once it is a few times larger than the other, hence it does not require many orders of magnitude fluctuation of the density contrast from galaxy to galaxy to accommodate the data. On the other hand, even though formation of a solitonic core is observed in simulations \cite{Schive:2014dra,Schive:2014hza,Mocz:2017wlg,Schwabe:2016rze}, the simulations have one single species with a fixed mass, done with DM only, and are verified only with a limited range of DM halos. For example, \cite{Schive:2014dra,Schive:2014hza} observe an empirical relation between the size of the solitonic core and the galaxy/halo size for the single scalar case, for galaxies with the size of $(10^9 \sim 5 \times 10^{11} ) M_\odot$. Neither the size of solitonic cores beyond this halo mass range, nor the physical mechanism that determines the size of the core size is currently known. It is conjectured that the size of the solitonic core could be due to the dynamical relaxation time being bottlenecked by the age of the universe \cite{Bar:2019pnz}. While we are not extrapolating the BEC size observed in \cite{Schive:2014dra,Schive:2014hza}, we leave an accurate dynamical estimate of the two BEC component $M_1/M_2$ for future work. 
%Dedicated **** two scalar, larger range, baryons
}

% In Fig.~*** we show that scanning over $\Phi_1(0)$ and $\Phi_2(0)$ we get a region in $C-M$ plane. Translating $C-M$ to $\rho_c - R_c$, one can see that the scatter region contains the line $\rho \sim 1/R^{1.3}$ that fits the data.

To summarize, we stress two points that distinguish two scalar BEC from the single scalar BEC at the galactic scale: 1) With a single scalar, {\hl the mass and radius of the BEC is fixed by the scalar potential instead of scalar dynamics. As a result,} it is highly nontrivial to find a potential that reproduces $\beta \approx 1.3$. This is verified by checking different combinations of potentials. This is no longer the case with two scalar BEC as we have a two-dimensional parameter space. {\hl In this case, both the scalar potential and the ratio of the two components (hence scalar dynamics) affect the BEC's mass-radius relation.} 2) In the single scalar BEC, it is even harder to accommodate the scattering of the points, even if one manages to find a potential that reproduces $\rho \sim 1/R^{1.3}$, other than attributing it to observational errors. In the two scalar scenario, it is natural to have a scattering due to dynamics that leads to a fluctuation of $\Phi_1(0)/\Phi_2(0)$ from the best fit curve. % {\color{red} 3) If a potential $V = m_1^2 \phi_1^2 + m_2^2 \phi_2^2$ is chosen, the galactic data requires a varying ratio of the two components. Given that  }

%**** Using the scan shown in the left of Fig.~\ref{fig:scan-of-phi1-phi2}, we can study the relation of $\Phi_1(0)$ and $\Phi_2(0)$.   

\section{Stellar Scale BEC Structure}
\label{sec:BosonStars}

Having discussed galactic scale BEC, we now turn to the properties of two-scalar stellar scale BECs.

Whether it is possible to stabilize a BEC system determines how dense the system can become before it is destroyed by self-gravity. In the single scalar scenario, there is only one way to support gravity to form $C\sim \mathcal{O}(0.1)$ dense objects:  by using repulsive $+\phi^{2n}$ potentials, whose model realizations have been shown to be non-trivial but possible \cite{Fan:2016rda}. Because of the presence of a second scalar, and non-gravitational interactions for both scalars and between them, we show that there are two new ways to support such systems to become dense enough. This could be relevant for gravitational wave signals from their binary mergers at LIGO-Virgo and LISA.

\subsection{Non-gravitational Interaction between Two Scalars}
\label{sec:non-grav-inter}

\begin{figure}[th]
  \centering
  \includegraphics[width=0.45\linewidth]{{Plots/mr_1_lam1_-1_lam2_-1_lam12_1}.pdf}
  \includegraphics[width=0.45\linewidth]{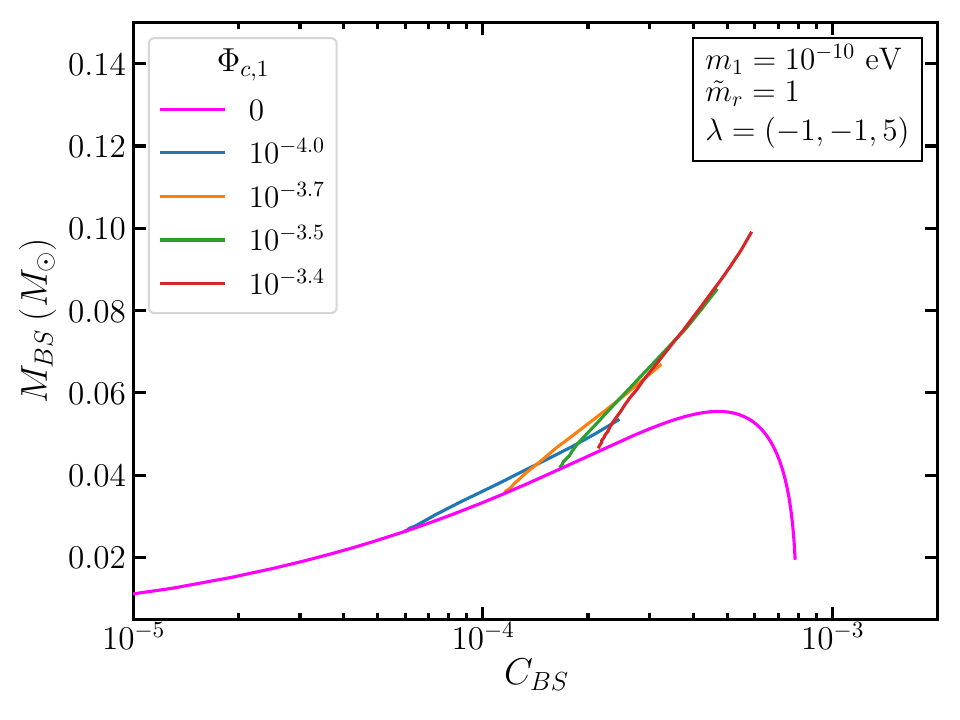}  
  \caption{The mass vs compactness for a scan over $\Phi_{2}(0)$ for various fixed $\Phi_{1}(0)$. $\tilde \lambda_{12}=1$ (left) is the same as right panel of Fig.~\ref{fig:HR-nonlinear}, for comparison with $\tilde \lambda_{12} = 5$ (right). The magenta curve corresponds to the single scalar limit. % The colored curves are achieved by fixing $\Phi_1(0)$ and increasing $\Phi_2(0)$.
    All curves are generated by scanning over $\Phi_2(0)$.
      Different colors correspond to fixing $\Phi_1(0)$ to different values.}  
    \label{fig:demo-of-lam12}
  \end{figure}

In Section~\ref{sec:non-relativistic}, we have already seen that the BEC structure can be drastically affected by the interaction $+\phi_1^2 \phi_2^2$ term. This has significance for exotic compact searches at stellar scale, such as binary mergers at LIGO and LISA. In this section, we show more details on the effect of non-gravitational interactions in the nonlinear regime, with different combinations of $\lambda_1, \lambda_2, \lambda_{12}$.
Among them, the most interesting case is $\lambda_1 <0, \lambda_2 <0, \lambda_{12}>0$. Without the non-gravitational interaction proportional to $\lambda_{12}$, neither $\phi_1$ or $\phi_2$ can form a BEC system that is compact enough to be detectable say at LIGO through gravitational wave radiated by the binary mergers. However, $+\phi_1^2 \phi_2^2$ provides pressure to support the gravitational collapse such that the system can be much denser as demonstrated in Fig.~\ref{fig:demo-of-lam12}. This is consistent with the analysis in Eq.~(\ref{eq:1}).

\begin{figure}[th]
  \centering
  \includegraphics[width=0.45\linewidth]{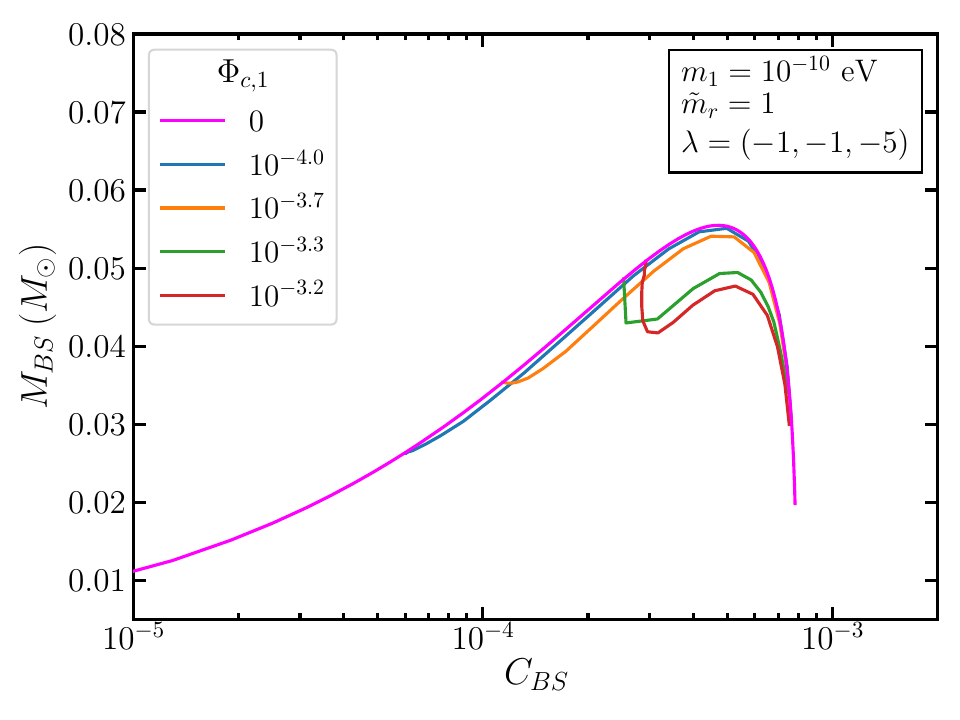}
  \includegraphics[width=0.45\linewidth]{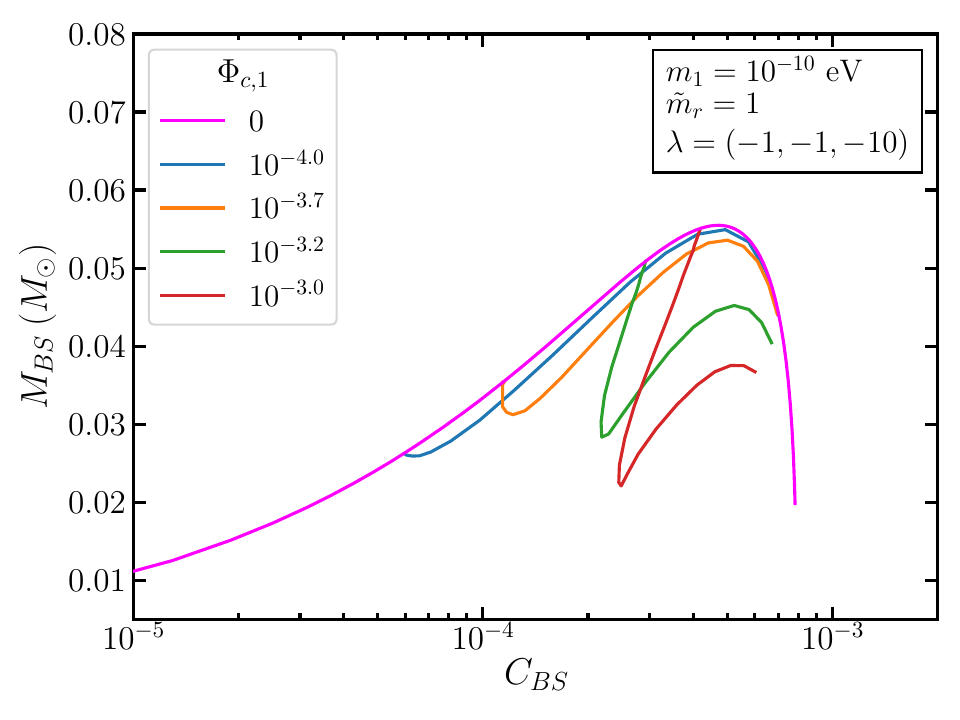}  
    \caption{When $\tilde \lambda_{12}=-5$ (left) and $\tilde \lambda_{12}=-10$ (right). One can see the extra interaction term can dominate over the kinetic term earlier. Once it happens, the Hamiltonian loses its local minimum and the system becomes unstable.     All curves are generated by scanning over $\Phi_2(0)$.
      Different colors correspond to fixing $\Phi_1(0)$ to different values.}  
    \label{fig:demo-of-lam12-negative}
  \end{figure}

In the case of $-\phi_1^2\phi^2_2$, we note that just as the non-gravitational self-interaction counterpart$-\phi^4$, it renders the system unstable once this term becomes important. This can also be understood using Eq.~(\ref{eq:1}). The effect is observed in the numerical solutions shown in Fig.~\ref{fig:demo-of-lam12-negative}. 

  {\hl
    We now comment on the difference between $\phi_1^2 \phi_2^2$ and $\phi^4$ types of interactions, and their impact on the resulted boson stars. One might think that if $\phi_1 \propto \phi_2$ everywhere, this two types of interactions are quite similar. However, this is rarely the case. At stellar scale, the formation history of BEC is highly nontrivial. In lack of a simulation, there is no reason to believe $\phi_1$, $\phi_2$ populate in every place proportionally. In the case of $V \sim -\phi_1^4 - \phi_2^4 + \phi_1^2 \phi_2^2$, for example, there could exist boson stars consisting mainly $\phi_1$ and those consisting mostly $\phi_2$ due to their separate fragmentation history \cite{Chen:2020cef}. Neither of the two types would be detectable because of the $-\phi^4$ potential. However, when the two stars merge, the interaction between the species now can support the self-gravity and allow the star to acquire more material to become denser, up to a point that it is detectable at LIGO/Virgo. In the case of  $V \sim +\phi_1^4 + \phi_2^4 - \phi_1^2 \phi_2^2$, on the other hand, the stars consisting mostly $\phi_1$ or $\phi_2$ could be very dense. However, unlike the single scalar BEC, the maximal compactness of either type is not capped by the metric fluctuation, but the term $-\phi_1^2 \phi_2^2$. When two boson stars of kinds merge, it is more likely to result in an unstable BEC due to the $-\phi_1^2 \phi_2^2$ term. In this case, one would not expect there remains a final boson star, but instead a phenomena dubbed Bosenova \cite{Arvanitaki:2009fg} could happen. % This is because even
    Even before the two finishes merging, the resulted BEC system goes beyond the critical maximal mass, due to $-\phi_1^2 \phi_2^2$. This could lead to gravitational wave different from those from single scalar $\phi^4$ BEC mergers \cite{Croon:2018ybs,Helfer:2018vtq}. As described in \cite{Chen:2020cef}, attractive self-interaction leads to fragmentation, it is only natural to expect an attractive interaction between the two scalars could also lead to fragmentation, but only when BEC's of different types encounter. This could lead to novel nonlinear behaviors in boson star formations.
    In both  scenarios above, we see the difference between $\phi_1^2 \phi_2^2$ and $\phi^4$ boson stars related to their formation history.

    On the aspect of observation, $\phi_1^2 \phi_2^2$ boson stars are also quite distinct from $\phi^4$ boson stars. From Eq.~(\ref{eq:1}), one can see that when $\phi_2^4$ term balances self-gravity, it leads to the usual mass curve, $M \propto C$. However, when $\phi_2^2 \phi_1^2$ balances self-gravity with $\phi_1$ being a sub-component, the mass curve is
    \begin{align}
      \label{eq:MvC-fix-M1}
      M \propto M_1^{1/3} C^{2/3}.      
    \end{align}
    When the sub-component $M_1$ is fixed, the mass curve is different from a $+\phi^4$ BEC system. On the other hand, due to the non-linear effect of gravity at small scales, one naturally expect the $\phi_1$ component, $M_1$ to vary from star to star. As a result, we expect some scattering around this mass curve, which is another feature that $\phi^4$ mass curve does not have.

    On the aspect of model building, as we will show in the following sections, ultralight scalars with interaction between multiple species can be achieved naturally. It is shown in \cite{Fan:2016rda} that it is nontrivial to build a $+\phi^4$ theory. Given that axions/ALPs all have a $-\phi^4$ interaction, we show that they can indeed lead to compact BEC structures if there are extra interactions - such as $+ \phi_1^2 \phi_2^2$ - that arises naturally from collective symmetry breaking and stabilizes the system, it could lead to dense axion stars. This hints for a different class of models compared to those that lead to a $+\phi^4$ theory. 
  }

\subsection{Implication for Gravitational Wave Detection}
\label{sec:implication-ligo}

% ** overlay LIGO band with $\lambda_1<0,\lambda_2<0,\lambda_{12}>0$ and curve reaching into $C>0.1$.
\begin{figure}[th]
  \centering
  \includegraphics[width=.6\textwidth]{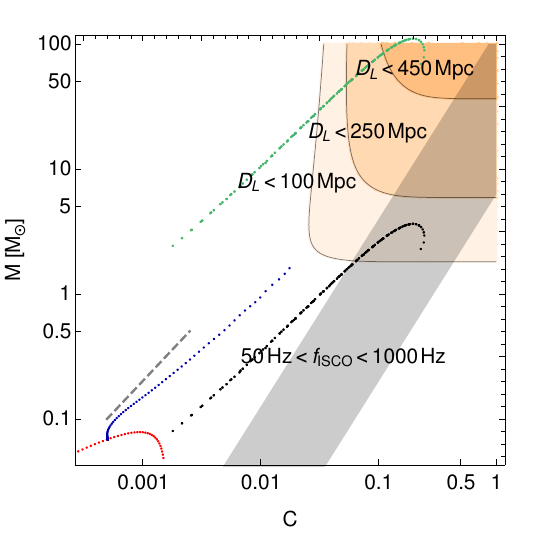}
  \caption{The orange bands correspond to the required $C-M$ region that can give SNR $\rho>8$ needed for detection, given different luminosity distances. The grey band corresponds to the region that $f_{ISCO}$ is in the LIGO sensitivity band, $50\;\mathrm{Hz}\sim 1000\;\mathrm{Hz}$. The red curve is the mass-compactness profile of a single axion with attractive $\phi_1^4$ with $\lambda_1 =-0.5$. The blue curve is achieved by adding a second axion to the red curve while fixing $\phi_1(0)$ and varying $\phi_2(0)$. The interactions are $\lambda_1 = \lambda_2 = -0.5, \lambda_{12} = +5$. {\hl We only show part of the curve due to computational complexity related to the stiffness in the equations.} % The grey curve is achieved by varying both $\phi_1(0)$ and $\phi_2(0)$, with the same interactions as the blue curve.
    % {\color{red} Describe $\lambda_2$. Is it $>0$ or $<0$? I suspect it's $>0$, which isn't what we are after.}
    The black and green curves are achieved by varying both $\phi_1(0)$ and $\phi_2(0)$, with the same interactions as the blue curve. The red and blue have mass set to $10^{-10} \;\mathrm{eV}$, the black $3\times 10^{-10} \;\mathrm{eV}$, and green $10^{-11} \;\mathrm{eV}$.
    % {  \color{red} comment on grey curve. comment on possible scattering}
{\hl    The grey dashed line is a guide for the eye, to show $M \propto C$, the mass relation one expects from a single scalar $+\phi^4$ BEC. One can see that indeed, when $\phi_1(0)$ is fixed, the mass curve has a slow smaller than one, as discussed in Eq.~(\pageref{eq:MvC-fix-M1}).}
}
  \label{fig:LIGO}
\end{figure}

In the past, LIGO-Virgo have observed numerous binary black hole mergers and a few neutron star mergers. Besides the tests of GR, there have been studies on probing new physics by detecting binary mergers consisting of exotic compact objects \cite{Giudice:2016zpa,Croon:2018ybs,Croon:2018ftb,Guo:2019sns,Bai:2018dxf} to name a few. While axions are well studied and motivated in particle physics, the compactness of axion stars is far below LIGO sensitivity, which renders a direct detection of axion star binary mergers impossible\footnote{It is noted that \cite{Braaten:2015eeu} argues that there exist a dense branch for axion stars, yet \cite{Visinelli:2017ooc} shows the lifetime of the dense branch is too short.}.
This changes when one takes into account extra scalars with non-gravitational interactions between them. As we have shown in Section~\ref{sec:non-grav-inter}, a repulsive $+\phi_1^2 \phi_2^2$ can support the system made of two axions up to $\mathcal{O}(0.2)$ compactness, a behavior that was only known to exist in repulsive self-interaction system \cite{Croon:2018ybs}.

For a generic binary system, the gravitational wave during its inspiral phase is
\begin{align}
  f_{GW}
  & = \sqrt{\frac{M_1 + M_2}{\pi^2 \ell^3} },
\end{align}
where $M_1, M_2$ are the masses of each inspiral object, $\ell$ the major semi-axis. Assuming equal mass system, the innermost stable circular orbit (ISCO) happens at $\ell = 6R$, with $R$ being the radius of the star. The gravitational wave frequency at ISCO can be expressed as
\begin{align}
  f_{ISCO}
  & =
    \sqrt{\frac{2M}{\pi^2 (6R)^3}}
    = \frac{C^{3/2}M_{Pl}^2}{2\pi \cdot 3^{3/2} M}
    \approx
    50\;\mathrm{Hz} \left (\frac{M_\odot}{M} \right ) \left (\frac{C}{0.04} \right )^{3/2}.
\end{align}
From the ISCO frequency one can extract the size of each object. The signal strength in frequency space is given as \cite{Khan:2015jqa}
\begin{align}
  \tilde h(f)
  &   =   \left ( \frac{\sqrt{5/24} G_N^{5/6}}{\pi^{2/3} c^{3/2}} \right )
  \frac{M_c^{5/6}}{f_{GW}^{7/6} D_L}
  \cr
  & \stackrel{{ISCO}}{\approx}
  7.2 \times 10^{-24} \mathrm{Hz}^{-1}
  \left (\frac{M}{M_\odot} \right )^2
  \left ( \frac{C}{0.04} \right )^{-7/4}
  \left ( \frac{D_L}{100\;\mathrm{Mpc}} \right )^{-1},
\end{align}
where the second line is estimated at $f_{GW} =  f_{ISCO}$. The signal-to-noise ratio (SNR) is computed as
\begin{align}
  \rho^2
  & =
    4 \int^{f_{ISCO}}_{0} \frac{|\tilde h(f)|^2}{S_n(f)} df,
\end{align}
where $S_n(f)$ is the detecor noise power spectral density \cite{LIGONoise:2018}. We require $\rho>8$ for a possible detection. The LIGO sensitivity band is shown in Fig.~\ref{fig:LIGO}, together with the mass-compactness profile of the BEC objects. It is observed that with a single axion-like particle ($-\phi^4$ self-interaction) the BEC structure is far from LIGO sensitivity band, while a repulsive interaction between the two scalars supports the system to $\mathcal{O}(0.2)$ region that is relevant at LIGO. 

With gravitational wave detectors LIGO-Virgo finishing O3 and being upgraded for higher sensitivity, KAGRA \cite{Akutsu:2020his} started the observation run, and LIGO-India planned to join the network in the near future, we emphasize that this serves as an example that the interferometry facilities have potential to probe fundamental particles and interesting interactions beyond gravity.

Intriguingly, LIGO-Virgo have recently detected a compact binary merger event (GW190521) with a total mass of $150\;M_\odot$ \cite{Abbott:2020tfl,Abbott:2020mjq}, with the primary mass $85^{+21}_{-14}\;M_\odot$ falling in the mass gap predicted by pair-instability supernova theory, $65 - 120 \; M_\odot$. In Fig.~\ref{fig:LIGO}, we show that a BEC system made of two scalars of mass $\sim 10^{-11} \;\mathrm{eV}$ can produce compact structures in the range of $100\;M_\odot$. While a dedicated analysis is needed to investigate its GW signals in our model, it is shown that certain types of boson stars could potentially reproduce the event \cite{CalderonBustillo:2020srq}.

\subsection{Comparison with previous work}
\label{sec:comp-with-prev}

Before we move on to the particle model, we briefly discuss the relation and differences with previous studies on multiple species BEC~\cite{Eby:2020eas}~and~\cite{Broadhurst:2018fei}.
In \cite{Eby:2020eas} the scenario of multiple axions are discussed in details. We list the main differences between this work and \cite{Eby:2020eas} and comment on them in more details.
\begin{itemize}
\item the analysis of \cite{Eby:2020eas} is performed in the non-relativistic limit, while we solve the full EKG system;
\item between different species \cite{Eby:2020eas} assume they only interact gravitationally, while we allow interactions $\propto |\phi_1|^2 |\phi_2|^2$;
\item analytical approximation to the solution of Sch\"odinger-Newton equation is carefully studied by \cite{Eby:2020eas} while we solve the EKG system both numerically and analytically.
\end{itemize}
We note that the first two points are relevant when the field value is large. % , \textit{i.e.} at stellar scales.
In particular, the first point captures the GR effect so that it allows us to apply the method to both the galactic scale BEC and the stellar scale. 
We also observe that in Eq.~(\ref{eq:Hnr}) our Hamiltonian precisely reproduces the result of \cite{Eby:2020eas} (Eq.~2.15 therein) in the limit $R_1 \approx R_2$. When $R_1 \neq R_2$, we have $\mathcal{O}(1)$ difference due to the choice of our anzats. We stress that we contain higher order terms $H_{metric}\propto G/R^3$, which have the origin of metric perturbation beyond Newtonian gravity, hence only shows up when one solves the EKG system. This is discussed in \cite{Croon:2018ybs} with a single species. 

Astrophysical implications of multiple axions are discussed in \cite{Broadhurst:2018fei}. We note a few differences compared to this work.
\begin{itemize}
\item The work of \cite{Broadhurst:2018fei} uses the so-called independent approximation, which treats solitons of different sizes separately. This allows them to go in to a regime where the mass of the heaviest one is  four orders of magnitude larger than the lightest one.
\item Similar to \cite{Eby:2020eas}, \cite{Broadhurst:2018fei} uses the Newtonian limit. 
\item \cite{Broadhurst:2018fei} assumes the scalars only interact gravitationally. 
\item The observational data include the Fornax Galaxy, central Milky Way; Ultra Faint Galaxies; and globular cluster 47 Tuc. In this work our fit at galaxy scale is motivated by \cite{Rodrigues:2017vto}.
\item \cite{Broadhurst:2018fei} also performs a Bayesian analysis to fit with the scalar mass, with the presence of extra nuisance parameters from the astrophysical environment.
\end{itemize}
%**tba**

\section{Model Realization}
\label{sec:model-realization}
In this section we discuss how  multiple light scalars with repulsive interactions between them can be realized from the point of view of particle physics model building. It is well known that light scalars can be generated by identifying them as pseudo-Nambu-Goldstone bosons (pNGB) after the spontaneous breaking of some approximate global symmetry. Applications of this idea  related to addressing the electroweak hierarchy include leveraging breaking patterns such as $SU(5) \rightarrow SO(5)$ \cite{ArkaniHamed:2002qy}, $SU(3)\times SU(3) \rightarrow SU(3)$ \cite{ArkaniHamed:2002qx}, $SU(6) \rightarrow Sp(6)$ \cite{Low:2002ws}, and $SO(6) \rightarrow SO(5)$ \cite{Schmaltz:2008vd}, to name a few. It was shown in \cite{Fan:2016rda} that similar collective symmetry breaking mechanisms can be used to generate repulsive self-interaction in the context of ultra-light dark matter, and potentially a large separation between the scalar mass and the symmetry breaking scale, which is needed to ensure the interaction between the scalars remains weak. We show that this can be extended to the two scalar scenario, which leads to a repulsive interaction between the two light scalars while being technically natural.

After spontaneous symmetry breaking of a global symmetry, such as $SU(6)$ to $Sp(6)$ in \cite{Low:2002ws}, one could end up with the following effective potential {\hl
\begin{align}
  \label{eq:2}
  c_1 f^2 \left |s + \frac{i}{2f} [\phi_2^\dagger \phi_1]  \right |^2+
    c_2 f^2 \left |s - \frac{i}{2f} [\phi_2^\dagger \phi_1]  \right |^2,
\end{align}
}
where $s, \phi_1, \phi_2$ are pNGB living in the quotient group, $SU(6)/Sp(6)$ in this example, and $c_{1,2}$ dimensionless coefficients that can be naturally small yet  positive. $s$ is a singlet under gauge transformations, while both $\phi_1$ and $\phi_2$ are gauge multiplets, doublets of two $SU(2)$'s in the specific case. $[ \phi \phi ]$ indicates a proper contraction with the gauge indices. One can see that each of the two terms in Eq.~(\ref{eq:2}) preserves a direction of the infinitesimal global transformation:
\begin{align}
  \phi_1 
  & \rightarrow \phi_1 + \epsilon_1, \cr
    \phi_2
  & \rightarrow \phi_2 + \epsilon_2, \cr
    s
  & \rightarrow s - \frac{i}{2f}( [\epsilon_2^\dagger \phi_1 ]+ [\phi_2^\dagger \epsilon_1] )
\end{align}
for the first term, and
\begin{align}
  \phi_1
  & \rightarrow \phi_1 + \epsilon_1, \cr
    \phi_2
  & \rightarrow \phi_2 + \epsilon_2, \cr
    s
  & \rightarrow s + \frac{i}{2f}( [\epsilon_2^\dagger \phi_1 ]+ [\phi_2^\dagger \epsilon_1] )
\end{align}
for the second term.
Equation~(\ref{eq:2}) generates a mass for the singlet $s$ to be $m_s^2 \sim f^2 (c_1 + c_2)$. 
Integrating out the scalar $s$, we have a interaction between $\phi_1$ and $\phi_2$ at tree level {\hl
\begin{align}
  V_{int}(\phi_1, \phi_2)
  \sim \frac{c_1 c_2}{c_1 + c_2} |[\phi_2^\dagger \phi_1]|^2. 
\end{align}
}
Mass terms for $\phi_1$ and $\phi_2$ are generated at one loop level,
\begin{align}
  V_{m} (\phi_1, \phi_2)
  \sim m_1^2 |\phi_1|^2 + m_2^2 |\phi_2|^2 + m_{12}^2 ([\phi_1^\dagger \phi_2] + h.c.),
\end{align}
where
\begin{align}
  m_1^2
  & \sim m_2^2 \sim \frac{c_1 c_2}{16\pi^2}f^2,
\end{align}
where we neglect the logarithmic part that is $\sim \mathcal{O}(1)$. Choosing a small $c_{1,2}$ leads to a large separation between the symmetry breaking scale and the scalar mass. {\hl The low energy potential is $V = V_m + V_{int}$. When $m_{12} \rightarrow 0$, the system has a total of four complex scalars that enjoy two separate $U(1)$'s, with generator $\theta_1$ and $\theta_2$: 
  \begin{align}
    \phi_1
    & \rightarrow \mathrm{e}^{+i \theta_1}\; \phi_1,
      \cr
    % \phi_1^d
    % & \rightarrow \mathrm{e}^{+i \theta_1}\; \phi_1^d,
    %   \cr
    \phi_2
    & \rightarrow \mathrm{e}^{+i \theta_2} \; \phi_2.
    %   \cr
    % \phi_2^d
    % & \rightarrow \mathrm{e}^{+i \theta_2} \; \phi_2^d,
  \end{align}
%  where $\phi_i^u$ ($\phi_i^d$ ) is the upper (lower) $SU(2)$ component of $\phi_i$. 
}
We hasten to add that the model we have proposed, which falls under the genre of ``Little Dark Matter" models introduced in \cite{Fan:2016rda}, is a proof-of-principle model construct of two-scalar ultralight scalar system with a repulsive interaction. %Further investigation would be required for a fully realistic treatment, which we leave for future work. 

\section{Conclusion}
\label{sec:Summary}
In this paper, we have demonstrated interesting features in a two scalar BEC system with spherical symmetry. We have  developed numerical code to solve the system exactly. We first verified its stability against radial perturbations by performing numerical time evolution. We then went on to  show two main features of the system:

1) \textbf{Galactic Scale:} The difference of the mass of the two scalars allows us to extend the BEC mass profile in the $C-M$ plane from a curve to a region, hence open up new parameter space. This is due to the fact that, with a fixed set of theory parameters, $[M_{BS}(\Phi_1(0)), C_{BS}(\Phi_1(0))]$ is extended to $[M_{BS}(\Phi_1(0), \Phi_2(0))$, $C_{BS}(\Phi_1(0), \Phi_2(0))]$. We show that this {\hl has interesting indications} to the problem \cite{Deng:2018jjz} that one scalar BEC cannot fit the dark matter core profile at the galactic scale. % Further studies taking into account the dynamics are left for future work.

2) \textbf{Stellar Scale:} At the stellar scale with $m\sim 10^{-10}\;\mathrm{eV}$, we show that the non-gravitational interactions between the two scalars $\Phi_1^2 \Phi_2^2$ can play an important role in stabilizing the system up to high compactness. This is similar to the fact that $+\phi^4$ self-interaction stabilizes the system and achieves compactness $C\sim \mathcal{O}(0.2)$. This has important implications for possible detections at LIGO-Virgo.
%
% 3) At the stellar scale, 
In addition, with $m_1\neq m_2$, we show that the transitioning from $\Phi_1$ dominating to $\Phi_2$ dominating holds for different choices of $\lambda's$. In particular, in the case of $\lambda_1\cdot \lambda_2<0$, the stability is determined by the dominating scalar with highest occupation number. % (\textit{i.e.} in the non-linear regime.)
This also hints at interesting phenomenology at gravitational detectors such as LIGO-Virgo and LISA.

Based on our results, there are several  interesting future directions. For example:  how does the presence of extra scalar(s) affect the cosmological evolution compared to the single scalar scenario? Given that it is known non-gravitational self-interaction can lead to an altered structure formation history as demonstrated in \cite{Fan:2016rda,Arvanitaki:2019rax}, the non-gravitational interaction between the two scalars will likely also change the structure formation process. This might change the bounds on ultra-light dark matter bounds derived from Lyman-alpha \cite{Irsic:2017yje,Nori:2018pka} and subhalo mass function \cite{Schutz:2020jox}. 

In addition, with two scalars the BEC spans a region in the mass-compactness plane instead of forming a curve. One could ask if all points in that region are equally possible to form. The answer is likely negative as galactic scale dynamics might affect the relation among the central density of each scalar. However, addressing this issue is beyond the scope of this work,  where our focus is on the mass profile of stable BEC systems. Indeed, the two questions described above may be related, and addressing them requires dedicated simulations. We leave this for future work.

\acknowledgments
We would like to thank Kfir Blum, Joshua Eby, JiJi Fan, Michael Geller, Mark Hertzberg, Hitoshi Murayama, Tim Tait, and Tomer Volansky for useful discussions at different stages of the work. 
We would also like to thank the two anonymous referees for their useful comments, which contribute to the improvement of the work. 
HG and KS are supported by DOE Grant desc0009956. CS is supported by the Foreign Postdoctoral Fellowship Program of the Israel Academy of Sciences and Humanities, partly by the European Research Council (ERC) under the EU Horizon 2020 Programme (ERC-CoG-2015 - Proposal n. 682676 LDMThExp), and partly by Israel Science Foundation (Grant No. 1302/19). JS acknowledges the REU program at the University of Oklahoma, which is supported by NSF grant 1659501.

\appendix{}

\section{Einstein Tensor}
\label{sec:Einstein-tensor}
We start with the metric as follows, 
\begin{align}
  g_{\mu\nu}
  & =
    \begin{bmatrix}
      -B(r) & 0 & 0 & 0 \\
      0 & A(r) & 0 & 0 \\
      0 & 0 & r^2 & 0 \\
      0 & 0 & 0 & r^2\sin^2\theta
    \end{bmatrix}.
\end{align}
We can then write out the non-zero Christoffel symbol components as
\begin{align}
  \label{eq:Christoffel}
  \Gamma_{t r}^t
  & =
    \frac{B'(r)}{2B(r)},
\qquad
  \Gamma_{tt}^r = \frac{B'(r)}{2A(r)},
    \cr
    \Gamma_{rr}^r
    & = \frac{A'(r)}{2A(r)}, 
\qquad 
      \Gamma_{\theta\theta}^r = -\frac{r}{A(r)},
      \qquad 
      \Gamma_{\phi\phi}^r = -\frac{r \sin^2(\theta)}{A(r)}, 
      \cr
      \Gamma_{r \theta}^\theta
  & = \frac{1}{r},
    \qquad\qquad
          \Gamma_{\phi\phi}^\theta = -\cos(\theta) \sin(\theta),
    \cr
    \Gamma_{r\phi}^\phi
  & = 
    \frac{1}{r},
    \qquad \qquad
          \Gamma_{\theta\phi}^\phi = \cot (\theta),
\end{align}
from which the Einstein tensor $G_\mu^\nu$ can be calculated
\begin{align}
G_t^t& =   -\frac{A'(r)}{r A(r)^2}+\frac{1}{r^2
       A(r)}-\frac{1}{r^2},
       \cr
       G_r^r
  & = \frac{B'(r)}{r A(r)
   B(r)}+\frac{1}{r^2
               A(r)}-\frac{1}{r^2},
    \cr
    G_\theta^\theta
    & = 
               -\frac{A'(r) B'(r)}{4
   A(r)^2 B(r)}-\frac{A'(r)}{2 r
   A(r)^2}+\frac{B''(r)}{2 A(r)
   B(r)}-\frac{B'(r)^2}{4 A(r)
   B(r)^2}+\frac{B'(r)}{2 r A(r)
      B(r)},
      \cr
      G_\phi^\phi
  & =
    -\frac{A'(r) B'(r)}{4 A(r)^2
   B(r)}-\frac{A'(r)}{2 r
   A(r)^2}+\frac{B''(r)}{2 A(r)
   B(r)}-\frac{B'(r)^2}{4 A(r)
   B(r)^2}+\frac{B'(r)}{2 r A(r) B(r)}.
\end{align}

\section{Numerical Procedure}
\label{sec:NumericalProcedure}
Typically the equations of motion are solved using the shooting method
which is successful in the one scalar case where the solution can
easily converge to the ground state configuration. The equations get
numerically difficult to solve when extra scalars are introduced,
particularly in the nonlinear regime where $\lambda$ can have
significant contribution to the total mass. A workaround was found by
implementing a relaxation algorithm into our personal code which
proved successful in solving the differential equations.

\subsection{Relaxation Method}
To find numerical solutions, we use the relaxation algorithm described in chapter 18 of \textit{Numerical Recipes} \cite{NumericalRecipes}. We first write the system in the standard form 
\begin{equation}
    \textbf{y}'(t)=\textbf{g}(t,\textbf{y}).
\end{equation}. 
We want to solve this system over the interval $[a,b]$. We start with a trial solution $\bar{\textbf{y}}$ that satisfies all boundary conditions. Then we choose a set of evenly spaced points $\{t_k\}_{k=0}^{M-1}$ spanning the interval. At each point except for $t_0$, we form the difference equations

\begin{equation}
    \textbf{E}_k=\bar{\textbf{y}}(t_k)-\bar{\textbf{y}}(t_{k-1})-(t_k-t_{k-1})\textbf{g}(t_\text{av},\bar{\textbf{y}}_{\text{av}}),
\end{equation}
where $t_\text{av}$ and $\bar{\textbf{y}}_{\text{av}}$ are the averages of $t_k$ and $t_{k-1}$, and $\bar{\textbf{y}}(t_k)$ and $\bar{\textbf{y}}(t_{k-1})$ respectively.
We want to adjust our trial solution so that each $\textbf{E}_k$ vanishes. Let $\Delta\bar{\textbf{y}}(t)$ represent the adjustments we need to make to the trial solutions at each of the grid points so that 
\begin{equation}
    \textbf{E}_k\big(\bar{\textbf{y}}(t_{k-1})+\Delta\bar{\textbf{y}}(t_{k-1}), \bar{\textbf{y}}(t_k)+\Delta\bar{\textbf{y}}(t_k)\big)=0.
\end{equation}
We can approximate $\Delta\bar{\textbf{y}}(t)$ at each of the grid points by expanding $\textbf{E}_k$ as a first-order Taylor series in $\Delta\bar{\textbf{y}}$. Then we have
\begin{align}
    \textbf{0} & =  \textbf{E}_k\big(\bar{\textbf{y}}(t_{k-1})+\Delta\bar{\textbf{y}}(t_{k-1}),\bar{\textbf{y}}(t_k)+\Delta\bar{\textbf{y}}(t_k)\big) \cr
    & \approx \textbf{E}_k\big(\bar{\textbf{y}}(t_{k-1}), \bar{\textbf{y}}(t_k)\big)+\sum_{n=0}^{N-1}\frac{\partial\textbf{E}_k}{\partial\bar{y}_n(t_{k-1})}\Delta\bar{y}_n(t_{k-1})+\sum_{n=0}^{N-1}\frac{\partial\textbf{E}_k}{\partial\bar{y}_n(t_{k})}\Delta\bar{y}_n(t_{k}),
\end{align}
where $N$ is the dimension of $\textbf{y}$ and $\bar{y}_n(t_k)$ is the $n$th component of $\bar{\textbf{y}}(t_k)$. Since we already know $\textbf{E}_k\big(\bar{\textbf{y}}(t_{k-1}),\bar{\textbf{y}}(t_k)\big)$, this gives us $N\cdot(M-1)$ equations for $N\cdot M$ unknowns. The remaining $N$ equations come from the boundary conditions.

These equations, together with the boundary conditions, allow us to solve for the first-order corrections $\Delta\bar{\textbf{y}}(t_k)$. By adding these corrections to $\bar{\textbf{y}}(t)$, we obtain a new trial solution. We then iteratively repeat this process with the new trial solution until the trial solutions converge. We determine convergence by measuring the average size of the components of the correction vectors $\Delta\bar{\textbf{y}}(t_k)$. Once the average size of the corrections becomes small enough, we assume that the trial solutions have converged to the correct solution.

\begin{figure}
    \centering
    \includegraphics[width=0.49\textwidth]{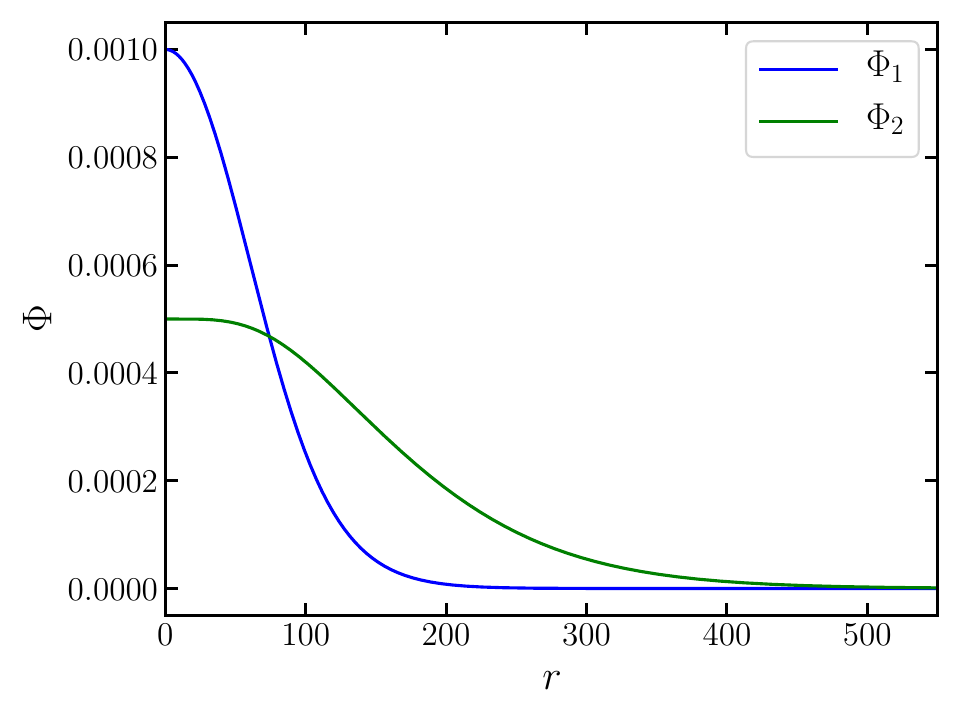}
    \includegraphics[width=0.49\textwidth]{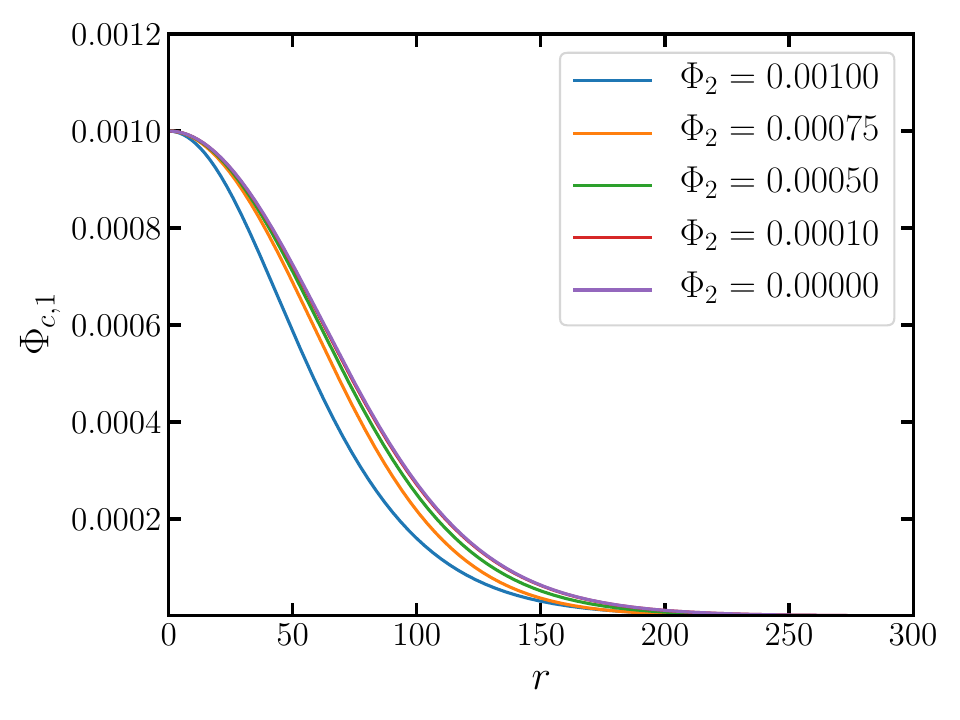}
    \caption{The left figure includes the wave forms for $\Phi_1$ and $\Phi_2$. The right figure fixes $\Phi_{1}(0) = 0.001$ and shows the effect of changing $\Phi_{2}(0)$. For $\Phi_{2}(0) \ll \Phi_{1}(0)$ the profiles behave like the one scalar case with $\Phi_{c} = 0.001$. When $\Phi_{1}(0) < \Phi_{2,1}$ we see a difference compared to the single boson case. Calculations were performed with $\lambda_1 = \lambda_2 = \lambda_{12} = 1$, $m_1 = 10^{-10}$ GeV,  and $\tilde{m}_r = 0.5.$}
    \label{fig:waveforms}
\end{figure}

\subsection{Static Case}
The profiles for $A,B,\Phi_1,\Phi_2,\Phi_1'$,and $\Phi_2'$ are found by solving the equations of motion using the relaxation method. The boundary conditions at the origin and at infinity are given by
\begin{align}
  \Phi_n(0)
  & = \Phi_{\text{c,n}} \\
  \Phi_n'(0)
  &= 0 \\
  A(0)
  &= 0 \\
  B(0)
  &= B_0 \\
  \lim_{r \rightarrow \infty}\Phi_n(r)
  &= 0 \\
  \lim_{r \rightarrow \infty}\Phi_n'(r)
  &= 0 \\
  \lim_{r \rightarrow \infty}B(r)
  &= \frac{1}{A(r)}.
\end{align}
For appropriate choices of the eigenvalues $\mu_{1,2}$ we can find the ground state configurations for $\Phi_{1,2}$. Although $\Phi_{1,2}$ must satisfy all of the boundaries conditions above, we can introduce constant differential equations to the equation of motion for the parameters of the problem without changing the physics of the system. This allows us to exploit the iterative process of the relaxation method to guess the values for $\mu_{1,2}$ until they converge to correct values as $r \rightarrow \infty$. We do this by including the following differential equations for $\mu_{1,2}$ into the relaxation method
\begin{equation}
    \frac{d \mu_n}{dr} = 0 \text{,   n =1,2}
\end{equation}
which allows in total 6 differential equations and 13 boundary conditions to be met. The numerical procedure to solve the equations of motion is as follows:
\begin{enumerate}
    \item Choose an initial guess for $A,B,\Phi_{1,2},\Phi_{1,2}^\prime$ that satisfies the boundary conditions.
	\item Run relaxation method on an interval $[0,r_{\text{out}}]$ \label{item:relaxation}
    \item If the error begins to diverge, recursively try a smaller interval and use that as an initial guess until it finds a solution.
    \item If $\Phi_{1,2} < 0$ or $\Phi_{1,2}^\prime > 0$, try again on a smaller interval with more grid points because an excited state was found.
    \item Check if $\Phi_{1,2}(r_{\text{out}}) < \epsilon$ where $\epsilon$ is a percentage of the initial central density. If true, $\Phi_{1,2}$ has decayed to its asymptotic value at $r_{\text{out}} = \infty$ and the ground state has been found. \label{item:converge}
    \item If condition \ref{item:converge} is not met increase $r_{\text{out}}$ and start from \ref{item:relaxation}.
\end{enumerate}

Once the ground state solutions are found for $\Phi_1$ and $\Phi_2$, the initial value of $B_0$ and the eigenvalues, $\mu_1$ and $\mu_2$ will be found to guarantee the boundary values are met. A sample plot of $\Phi_{1,2}$ is included in Fig.~\ref{fig:waveforms}. The left figure is a representative plot of the wave profiles for $\Phi_{1}$ and $\Phi_2$. The equations of motion couple both scalars together.  We can see from the right figure the impact the second scalar has on $\Phi_1$ by varying its central density. For $m_{1} = 10^{-10}$ eV and $\tilde{m}_r = 0.5$, the equation of motions look like the single scalar case when $\Phi_{2}(0) \ll \Phi_{1}(0)$. However, when $\Phi_{2}(0) \sim \Phi_{1}(0)$, the wave profile of $\Phi_{1}$ deviates from the single scalar case as expected.   

\subsection{Time Evolution}
\label{sec:time-evolution}
To ensure the stability of the equations of motion, we must first check how the equations of motion evolve under small radial perturbations. We first approximate the partial derivatives in the equations motion as central finite differences :

\begin{equation}
    \frac{\partial f}{\partial r} = \frac{ f_{j+1}^{i}(r,t) - f_{j-1}^{i}(r,t) }{2 \Delta r},
\end{equation}
\begin{equation}
    \frac{\partial f}{\partial t} = \frac{ f_{j}^{i+1}(r,t) - f_{j}^{i}(r,t) }{ \Delta t},
\end{equation}
\begin{equation}
    \frac{\partial^2 f }{\partial r^2} = \frac{f_{j+1}^i(r,t) - 2 f_j^i(r,t) + f_{j-1}^i (r,t)}{\Delta r ^2},
\end{equation}
\begin{equation}
    \frac{\partial^2 f }{\partial t^2} = \frac{f_{j}^{i+1}(r,t) - 2 f_j^i(r,t) + f_{j}^{i-1} (r,t)}{\Delta t ^2},
\end{equation}
where $i$ and $j$ correspond to steps in space and time respectively. The step sizes are given by $\Delta r$ and $\Delta t$. We note that these expressions are only valid for $i \in [1,N_t-1]$ and $j \in [1,N_r-1]$ where $i = 0,...,N_t$ and $j = 0,...,N_r$. For the endpoints we use either backwards or forward difference. Using finite differences we see that the two Klein-Gordon equations of motion give

\begin{equation}
    \Phi_{j,(1,2)}^{i+1} = 2 \Phi_{j,(1,2)}^i - \Phi_{j,(1,2)}^{i-1} + \Delta t^2 F\left(r,A,B,\Phi_1,\Phi_2,,A^\prime, B^\prime, \Phi_1^\prime,\Phi_2^\prime \right).
    \label{eq:timestep}
\end{equation}

To find the time evolution of the system we perform the following steps:

\begin{enumerate}
    \item Solve the static equations of motion.
    \item Perturb $\Phi_{1,2}$ by a factor of $(1 + \epsilon)$. 
    \item Perform the first time step in the Klein-Gordon equations for $\Phi_{1,2}$ using Eq~\ref{eq:timestep} where the static solution is $\Phi_{j,(1,2)}^{i-1}$ and the perturbed solution is $\Phi_{j,(1,2)}^{i}$. 
    \item Perturb $\Phi_{1,2}$ by a factor of $(1 + \epsilon)$.
    \item Solve the remaining two differential equations using the Relaxation Method to get $A_j^{i+1}$ and $B_j^{i+1}$.
    \item Repeat. 
\end{enumerate}
Sufficient time steps were performed following the above procedure to ensure the stability of the time evolution equations for sample benchmark points. 

The physical time, $t$, used in the equations of motion is in units of $m^{-1}$ with the corresponding dimensionless time, $\tilde{t}$, given by 
\begin{equation}
	t = \frac{1}{m} \tilde{t}.	
\end{equation}
The time evolution of the single scalar case was previously studied in \cite{Schiappacasse:2017ham} for both the stable and unstable branch with different re-scaled quantities. To match with the notation there, we compare our dimensionless variable with the ones used in \cite{Schiappacasse:2017ham}. The potential used in the analysis considers the non-relativistic interaction term  
\begin{equation}
	V_{nr}  = \frac{\psi^{*2} \psi^2}{16 f^2}
\end{equation}
with the following ansatz for the scalar field
\begin{equation}
	\Phi = \frac{1}{\sqrt m} \Psi(r) = \sqrt{\frac{N}{\pi m R^3}}e^{-r/R}
\end{equation}
where $R$ is the decay length scale and $N$ is the total number of particles.  The re-scaled quantities are 
\begin{align*}
	R &= \frac{M_{Pl}}{m f} \hat{R}  \\
	N &= \frac{M_{Pl} f}{m^2} \hat{N} \\
	\Psi &= \frac{\sqrt m f^2}{M_{Pl}}\hat{\Psi},
\end{align*}
where $\hat X$ is the dimensionless counterpart of variable $X$ used in \cite{Schiappacasse:2017ham}.
The corresponding time evolution equation of the scalar field in dimensionless units is 
\begin{equation}
	i \frac{\partial \hat{\Psi}}{\partial \hat{t}} = - \frac{1}{2 \hat{r}} \frac{\partial^2 }{\partial \hat{r}^2} \left( \hat{r} \hat{\Psi} \right) + \hat{\phi}_N \hat{\Psi} - \frac{1}{8} |\hat{\Psi}|^2 \hat{\Psi}
\end{equation}
where $\hat \phi_N$ is the newtonian potential. If we restore the physical parameters in the above equation, the physical time will become  
\begin{equation}
	t = \frac{M_{Pl}^2}{m f^2} \; \hat{t}.
\end{equation}
This allows for simple comparison by setting the physical times equal to each other. %  The scalar field has different dimensions therefore we can compare the two different definitions by comparing the interaction terms. After plugging in the definitions for the re-scaled quantities, the relationship between our definition and the one used in \cite{Schiappacasse:2017ham} for the scalar field becomes
% \begin{equation}
% 	\tilde{\phi} = \left( \frac{-\pi}{\tilde{\lambda}} \right)^{1/4} f^{3/2} G^{3/4} \tilde{\psi}
% \end{equation}  
% which is dimensionless when relating $G$ to Planck's constant.
The scalar field is related by
\begin{align}
  \tilde \Phi
  & =
    \frac{\sqrt{4\pi} f^2}{M_{Pl}^2} \hat \Psi.
\end{align}

\section{Single Scalar Limit}
\label{sec:single-scalar-limit}
\begin{figure}
    \centering
    \includegraphics[width=0.49\textwidth]{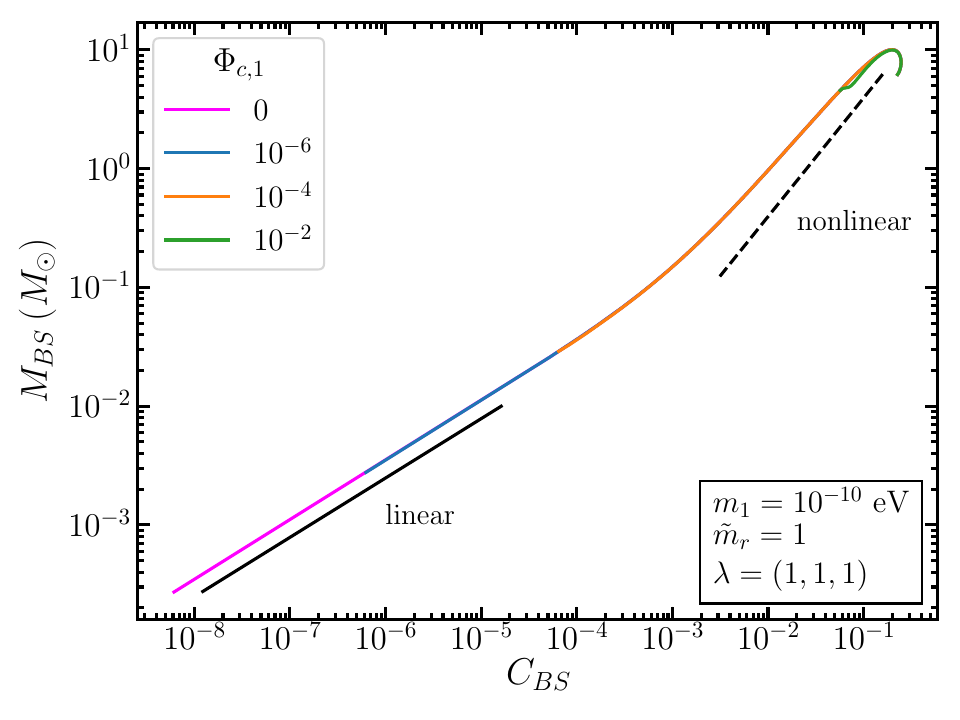}
    \includegraphics[width=0.49\textwidth]{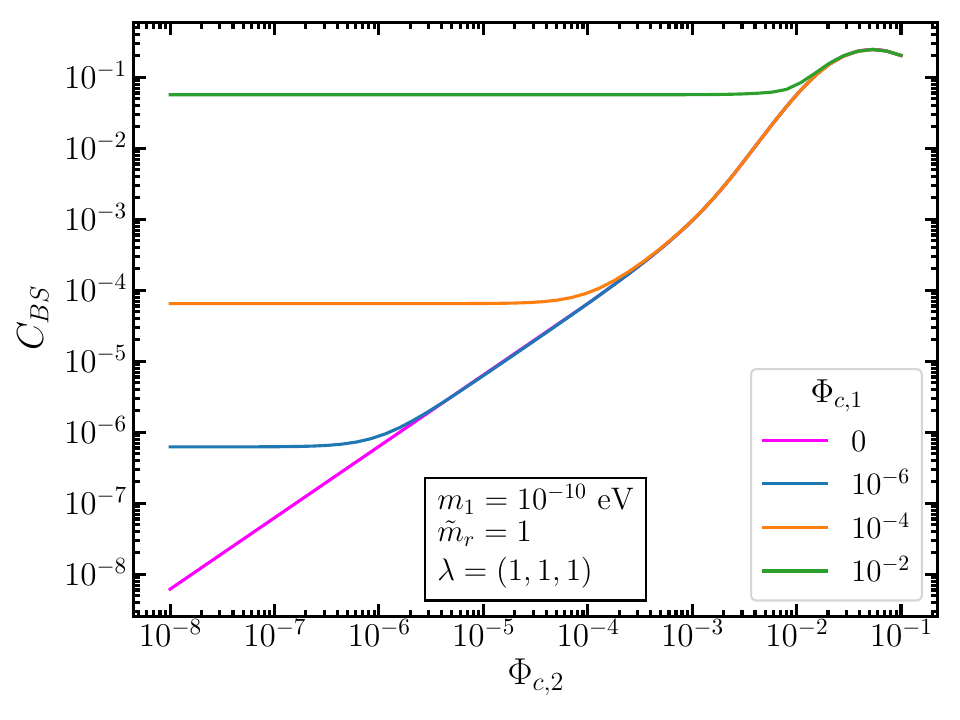}
    \caption{ {BEC structure obtained with equal mass for the two scalars and different fixed central density of one scalar, $\Phi_{1,c}$, and different coupling choices.} Left: the total mass vs compactness for various $\Phi_{1}(0)$ and $\Phi_{2}(0)$ for $\lambda_1 = 1$, $\lambda_2 = 1$, and $\lambda_{12} = 1$. The solid lines correspond to fixed $\Phi_{1}(0)$ while scanning over $\Phi_{2}(0)$.  The solid magenta curve corresponds to the single scalar limit by setting $\Phi_{1}(0) = 0$ and scanning over $\Phi_{2}(0)$. The solid and dashed black lines represent the linear and nonlinear scaling cases when $M_{BS}$ is derived from Eq.~\ref{eq:Hnr}. Right: the compactness versus $\Phi_{2}(0)$ for different values of $\Phi_{1}(0)$. It plateaus to a fixed value of $C_{BS}$ when $\Phi_{2}(0)$ is small and the star is dominated by $\Phi_{1}$.{The turning point occurs 
    when $\Phi_{2}(0) \approx \Phi_{1}(0)$}}
    \label{fig:cvm-mr_1_lam_1_lam2_1_lam12_1}
  \end{figure}

\begin{figure}
  \centering
    \includegraphics[width=0.6\textwidth]{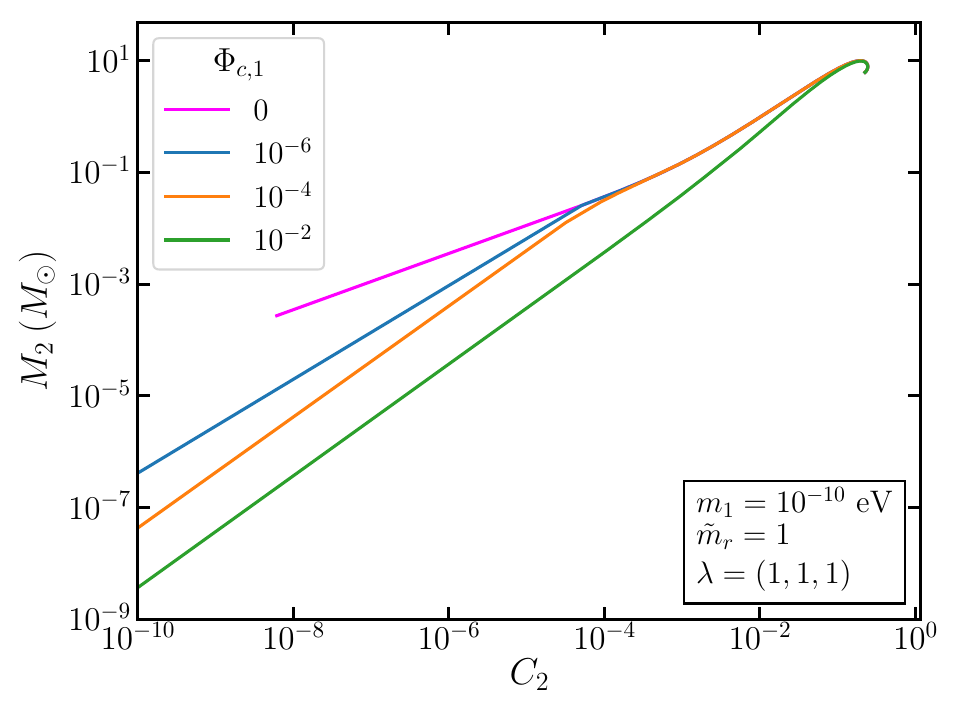}
  \caption{{The changes to a single scalar BEC profile due to the existence of another scalar
with varying central density.} The mass versus compactness only taking into consideration the contribution from $\Phi_{2}$. The single scalar limit is given by the blue curve. The other curves at scanning over $\Phi_{2}(0)$ at different fixed values of $\Phi_{1}(0)$.}
  \label{fig:singlescalarlimit}
\end{figure}

In this section we verify that in the limit of $m_1 =m_2$, and $\lambda_1 \sim \lambda_2$, one recovers the single scalar limit as expected. 
In the non-relativistic section it was stated that the scalar limit occurs when one of the number densities dominates over the other. The number density in the non-relativistic limit is given by 
\begin{equation}
  N_i = 4\pi \int dr r^2 m_i |\Phi_i(r)|^2
\end{equation}
{where the index $i=1,2$ corresponds to each scalar,} and the integral is over the scalars central density. This is related to the mass of the star for the two scalar system as 
\begin{equation}
  M = m_1 N_1 + m_2 N_2.
\end{equation}
The single scalar limit is taken for when $\Phi_{2}(0) \gg \Phi_{1}(0)$ and vice versa.
In Fig.~\ref{fig:cvm-mr_1_lam_1_lam2_1_lam12_1} we show that the single scalar limit can be recovered when $\Phi_1(0)$ is chosen to be small. 
%In Fig.~\ref{fig:singlescalarlimit} we show the contribution to the Boson star mass when we neglect the first scalar in Eq.~\ref{eq:Mbs} and scan over $\Phi_{2}(0)$ for different values of fixed $\Phi_{1}(0)$.
In Fig.~\ref{fig:singlescalarlimit}, we show $C_2$ versus $M_2$ for the second scalar's contribution % (\ghk{How to define $C_2$ and $M_2$. Is $M_2$ the mass of $\Phi_2$ within a fixed boson star radius(call it $R_{BS}$)?
% Is $C_2=M_2/R_{BS}$? Is $R_{BS}$ defined w.r.t the entire boson star, and not just $\Phi_2$?
% })
to the BEC system  mass and compactness where we scan over $\Phi_{2}(0)$ for different values of fixed $\Phi_{1}(0)$. %\textcolor{red}{(fix figure)} 
% The blue curve represents the total mass of the star in the single scalar limit.
The curves represent scans over the two central densities. Each curve for the two scalar scans begin with $\Phi_{2}(0) \ll \Phi_{1}(0)$ on the far left points. The curves all begin with masses much below the single scalar limit which means that the second scalar contribuion to the total mass of the star is subdominate and we can safely assume $N_2 \ll N_1$.  Each one of the curves eventually lead towards the single scalar limit when $\Phi_{2}(0)$ grows larger. This behavior confirms the analytical approximationa of the single scalar limits to determine the nonlinear and linear regimes done in section \ref{sec:non-relativistic}.
% (\ghk{If we draw a horizon line, then it will intersect with the single scalar curve(blue) and one of the 
% fixed $\Phi_{1}(0)$ line. For these 2 fixed points, the one in the presence of $\Phi_2$ is more
% compact. Does this mean $\Phi_{2}$ is more compact when $\Phi_1$ is present? Of course, this can also be 
% caused by a increased $\Phi_{2,c}$ and a reduced $R_{BS}$(defined by the whole BS rather than just $\Phi_2$).
% }) 


\section{Transitioning from $\Phi_1$ to $\Phi_2$ in the Nonlinear Regime}
\label{sec:non-grav-self}
%{\color{red} *** put section into the appendix ***}

{\hl We numerically verify that in the nonlinear regime, the transition between $\phi_1$ dominating to $\phi_2$ dominating still happens as expected.} As seen in Section~\ref{sec:non-relativistic}, a two scalar hierarchy $m_1 < m_2$ interpolates two scenarios where $\Phi_1$ dominates the system ($N_1 \gg N_2$,) and that where $\Phi_2$ dominates ($N_2 \gg N_1$.) We verify that this transitioning behavior still persists in the nonlinear regime. 
\begin{figure}
    \centering
    \includegraphics[width=0.49\textwidth]{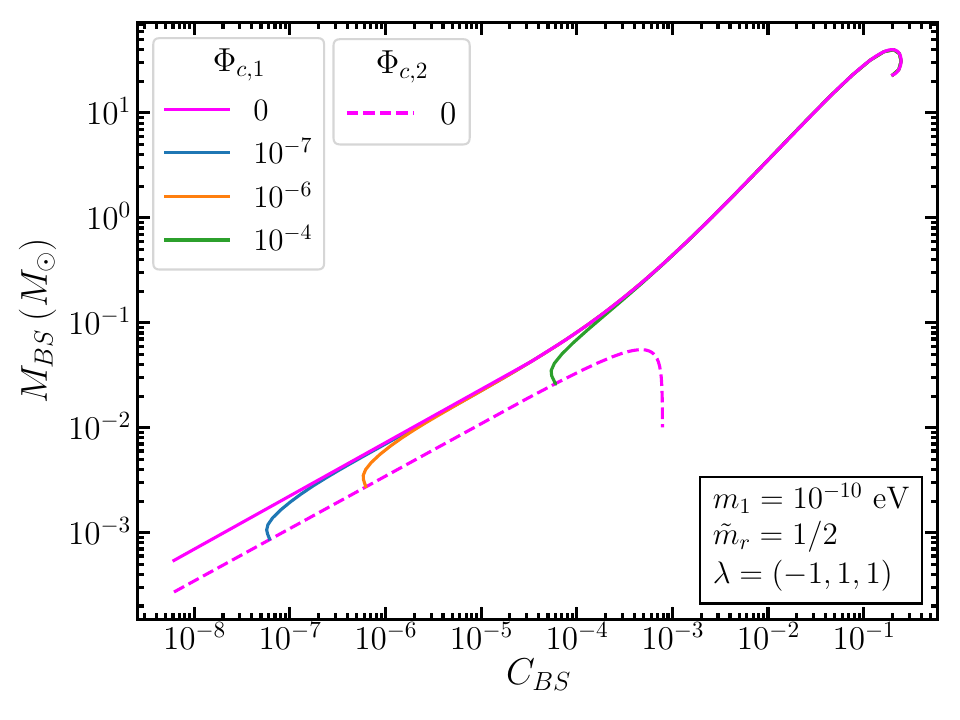}
    \includegraphics[width=0.49\textwidth]{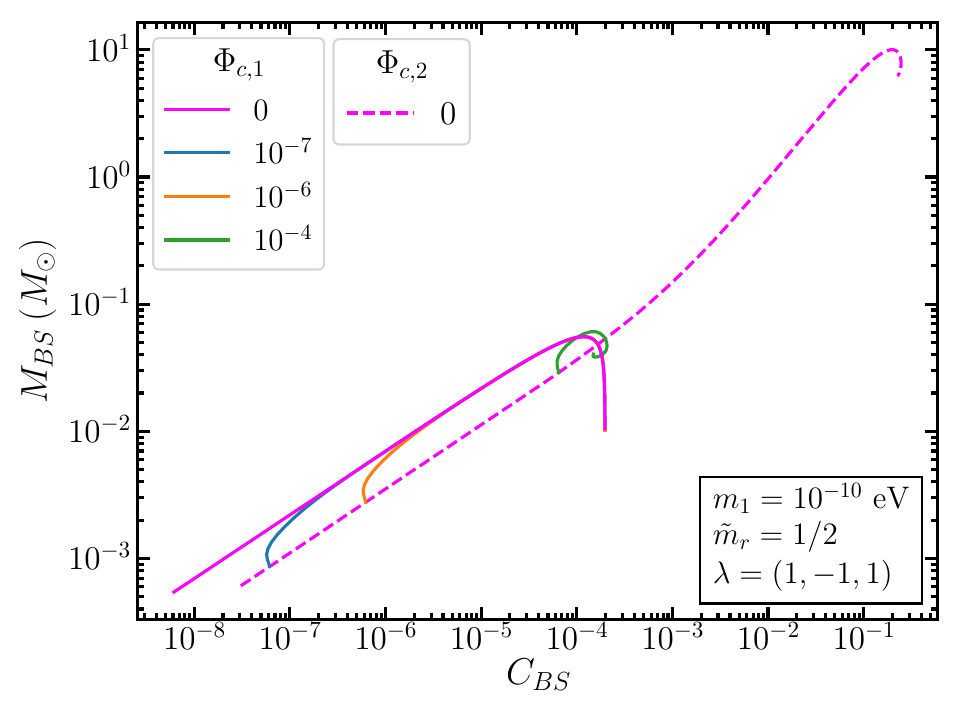}
    \caption{The total mass vs compactness for various values of $\Phi_{1}(0)$ and $\Phi_{2}(0)$, with $\lambda_1 = -1$, $\lambda_2 = 1$ (left) and  $\lambda_1 = 1$, $\lambda_2 = -1$ (right). All solid curves are generated by scanning over $\Phi_2(0)$ and fixing $\Phi_1(0)$ at labeled values, while the dashed curve is setting $\Phi_2(0)$ to zero and scanning over $\Phi_1(0)$.
% In both plots, if $\Phi_2$ had not existed, the $\Phi_1$ alone would have given the dashed curve. 
% Similarly, if $\Phi_1$ had not existed, $\Phi_2$ alone would have  given the solid magenta curve. 
% Fixing $\Phi_1(0)$ at different values and scanning over $\Phi_2(0)$, we essentially let the system transition from $\Phi_1$ dominant to $\Phi_2$ dominant at different places (the other colored curves.) 
}
    \label{fig:cvm-color-lams_neg}
  \end{figure}
 
\begin{figure}
	\centering
        \includegraphics[width=0.49\textwidth]{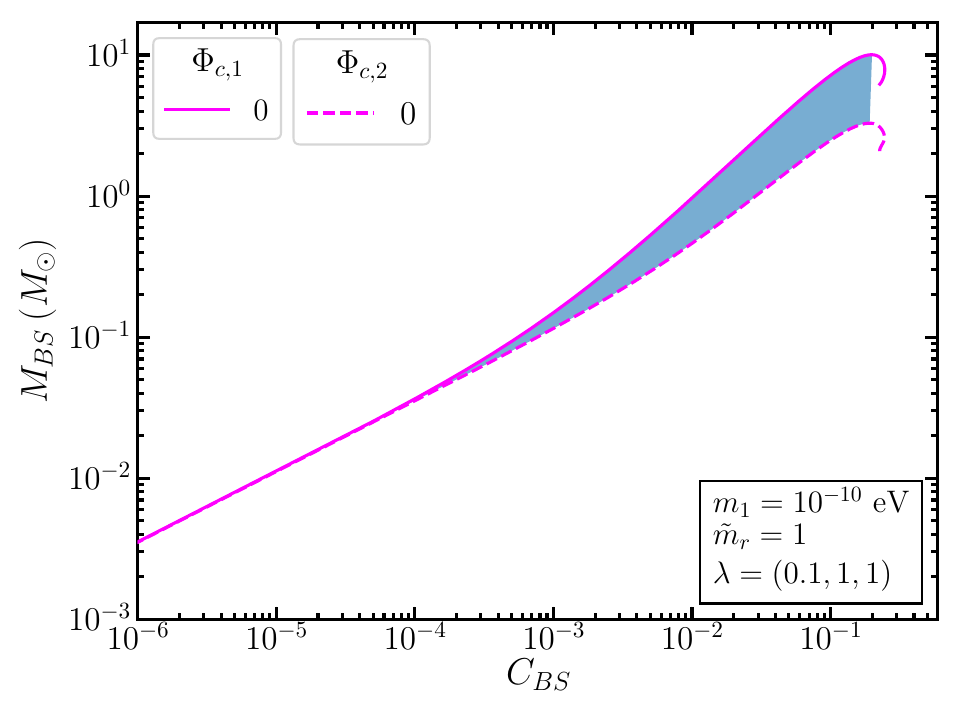}
        \includegraphics[width=0.49\textwidth]{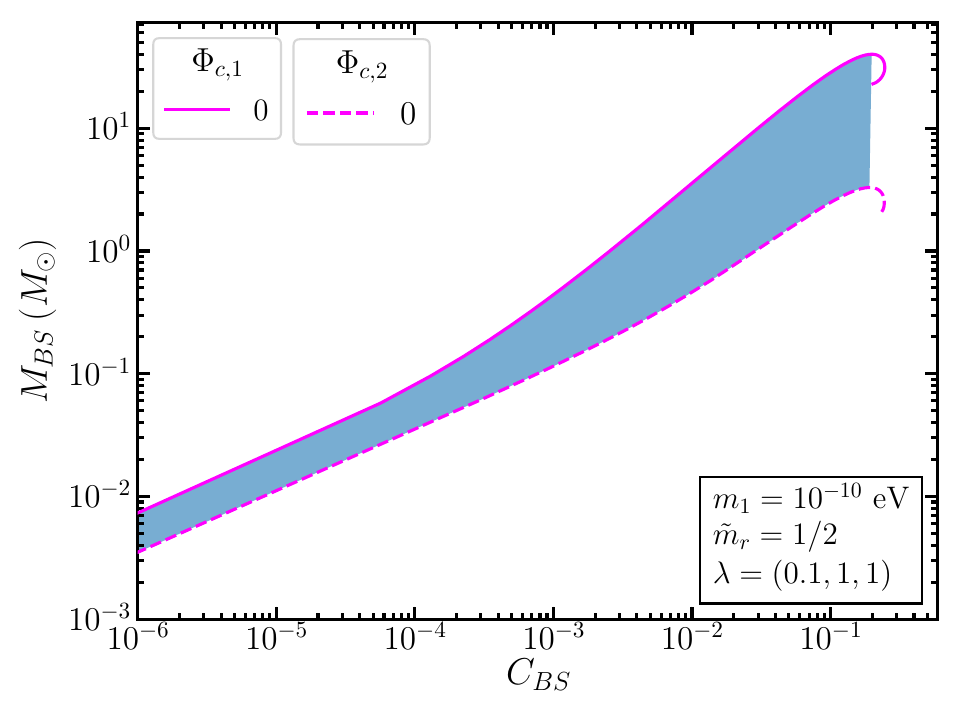}
	\caption{The mass profile of the BEC with $\lambda_1 = 0.1$, $\lambda_2 = 1$, and $\lambda_{12} = 1$.  The mass ratio between the two scalars is $\tilde{m}_r = 1$ in the left figure and $\tilde{m}_r = 1/2$ in the right figure.  The shaded region is the region a stable BEC system can form.}
	\label{fig:cvm-filled}
\end{figure}

  One observes that when one scalar has a stable nonlinear self-interaction (e.g. $+\phi^4$,) and the other unstable self-interaction (e.g. $-\phi^4$,) once the system transitions from the unstable scalar dominating to the stable scalar dominating, it is then stabilized, and vice versa. This can be see in Fig.~\ref{fig:cvm-color-lams_neg}. Another way of seeing this transitioning effect is through a less drastic setup, with $\lambda_1, \lambda_2>0$ but have different values. The $C-M$ curve has different shape if $\Phi_1$ or $\Phi_2$ forms BEC alone. In the two scalar system, by arranging $\Phi_1(0)$ and $\Phi_2(0)$ carefully, one can get any point in between the two curves shown as the shaded region in Fig.~\ref{fig:cvm-filled}.

\bibliographystyle{unsrt}
\bibliography{bib}
\end{document}